# Time Evolution of a Supply Chain Network: Kinetic Modeling

by

[a,b]Biswajit Debnath, [c]Rihab El-Hassani, , [a,*]Amit K Chattopadhyay, [d]T Krishna Kumar, [e]Sadhan K Ghosh, [f]Rahul Baidya

[a]Aston University, Department of Mathematics, ASTUTE, Birmingham B4 7ET, UK

[b] Chemical Engineering Department, Jadavpur University, Kolkata 700032, India

[c]ENSIIE – Ecole Nationale Suprieure d'Informatique pour l'Industrie et l'Enterprise, Paris, Evry, France

[d]Rockville-Analytics, Rockville, Maryland 20850, USA

[e]Department of Mechanical Engineering, Jadavpur University, Kolkata 700032, India

[f]Institute of Engineering and Management, Management House, Salt Lake 700091, Kolkata

[*]*Corresponding Author:* a.k.chattopadhyay@aston.ac.uk

**Abstract**

Resilient supply chains are often inherently dependent on the nature of their complex interconnected networks that are simultaneously multi-dimensional and multi-layered. This article presents a Supply Chain Network (SCN) model that can be used to regulate downstream relationships towards a sustainable SME using a 4-component cost function structure - Environmental (E), Demand (D), Economic (E), and Social (S). As a major generalization to the existing practice of using phenomenological interrelationships between the EDES cost kernels, we propose a complementary time varying model of a cost function, based on Lagrangian mechanics (incorporating SCN constraints through Lagrange multipliers), to analyze the time evolution of the SCN variables to interpret the competition between economic inertia and market potential. Multicriteria decision making, based on an Analytic Hierarchy Process (AHP), ranks performance quality, identifying key business decision makers. The model is first solved numerically and then validated against real data pertaining to two Small and Medium Enterprises (SMEs) from diverse domains, establishing the domain-independent nature of the model. The results quantify how increases in a production line without appropriate consideration of market volatility can lead to bankruptcy, and how high transportation cost together with increased production may lead to a break-even state. The model also predicts the time it takes a policy change to reinvigorate sales, thereby forecasting best practice operational procedure that ensures holistic sustainability on all four sustainability fronts.

# 1. Introduction

The term "Supply Chain" was coined by Banbury in 1975 to analyze supply networks across multiple functions and organizations. The concept has evolved over the years, both from the perspective of the American Production and Inventory Control Society (APICS) dictionary [8], as also from The Supply Chain Council [23]. A modern Supply Chain Network (SCN) is a complex multi-echelon, multi-stakeholder and multi-parameter network that includes uncertainty at various stages accompanied by risk management in decision making [9]. From a Life Cycle Assessment (LCA) perspective, a SCN can be based on the approach taken, such as cradle-to-cradle, cradle-to-gate, gate-to-gate etc. [7,31].

This article studies a generalized supply chain model that can analyze interrelationships between environmental, demand, economics and social logistics to identify a risk assessed generic supply chain, based on a cradle-to-utilization model. Figure 1 below outlines a flowchart explaining the SCN within a 3-step SCN that includes a supply side connecting materials and component manufacturers as input in step 1, internal operations involving process technologies involved in material processing, manufacturing and further quality assurance in step 2, and finally the critical demand aspect for business sustenance, consisting of users and logistics providers, leading to the output in step 3.

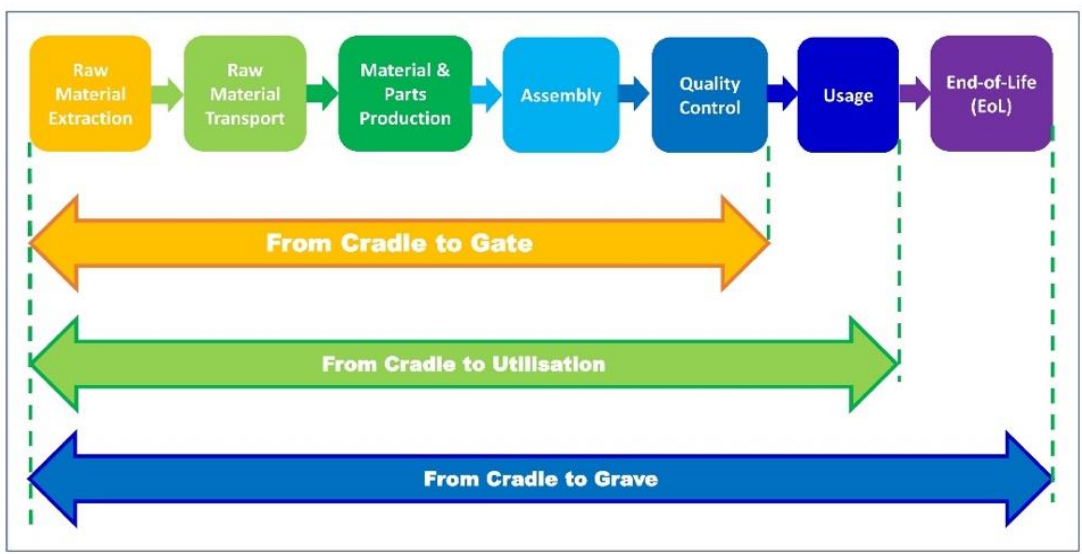

**Figure 1**: Boundaries of Supply Chain based on LCA

The first objective is to relate these three production processes and the four cost kernels (Figure 1) to various production process indicators (seven). Interrelationships between these seven elements can be defined both by internal knowledge uncertainty and external uncertainty that are translated into risk accompanying the four cost kernels. The sustainable strategies are then derived from constrained risk minimization [14]. The efficiency of a supply chain in adapting to these uncertainties plays a key role in achieving business success, so that even in situations with large uncertainty, as with the ongoing COVID situation, internal management can identify "performance windows" within which the supply line can still be sustained or at least reoriented. The production systems thus need to be designed in such a way that they combine high productivity with economic efficiency, while safeguarding sustainability of the business under the threat of business uncertainty.

The concept of Sustainable Production (SP) is thus strategically linked with 'greening' the internal operations of a Supply Chain Network (SCN). Note that while 'greening' explicitly addresses the environment kernel in the 4-pillar setup, the interrelationships with the other three modes ensure a 'holistic greening' sustenance of the model. Traditionally SP is based on the three basic pillars of sustainability i.e., environmental, economic, and social [5]. As the Brundtland report shows, the pillars of sustainability can be logically customized to four or more pillars conforming to the evolving definition of sustainability [33]. The other pillars can be institutional [10], operational [11], human [27] or even cultural [36]. In this study, our focus is strictly on supply chain sustainability, based on a 4-dimensional sustainability kernel, combining environmental, economic, social and demand uncertainty components.

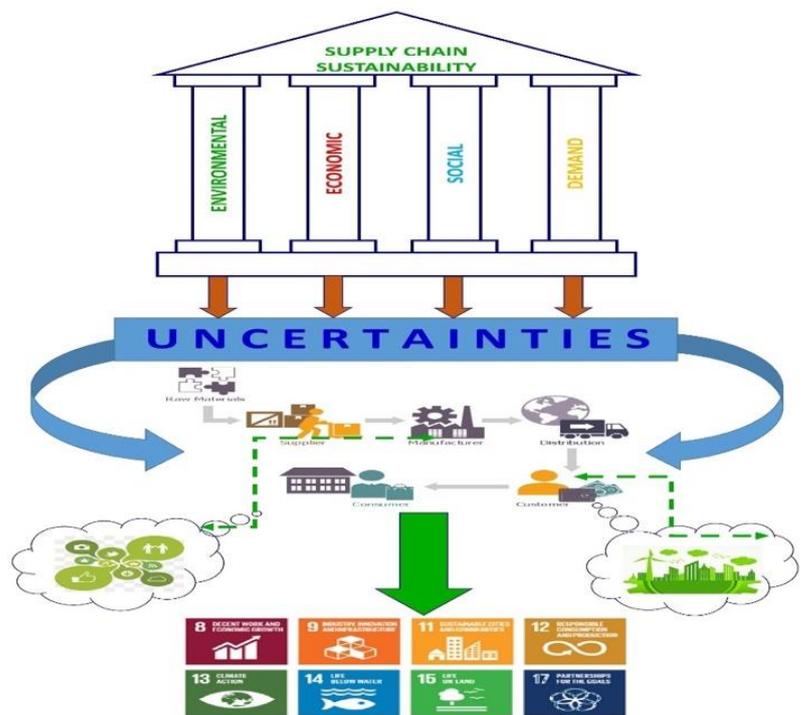

Demand plays a crucial role in the supply chain, and even the slightest market volatility affects the demand which in turn affects the sustainability of the supply chain. According to [23]. The four identified SCN pillars have several minute operational nodes which can pose serious threat to the supply chain as well as to the business logistics. The identified uncertainties are directly linked to the sustainability pillars, as there exists an inverse relationship between the uncertainty parameters and the sustainability pillars. Lower the uncertainty, higher is the sustainability (Figure 2).

**Figure 2**: Interrelationship between Supply Chain Uncertainties, Green Strategies and SDGs

The concept of sustainable SCN is intertwined with these four pillars, offering enough scope to incorporate the appropriate elements of SP (Figure 2). The agenda then is to develop best prevention strategies that can reduce uncertainty braided risk in devising SP strategies.

In Figure 2, the pillars show the supply chain sustainability dimensions, the brown arrows signify the contribution of the four pillars towards uncertainties, the blue arrows signify uncertainties affecting SCN, cloud callouts linked with double-sided arrows suggest that connection of green strategies is a feedback process. The

Green Arrow links the production process to sustainable output leading towards addressing SDGs 8,9,11-15 and 17 (https://sdgs.un.org/goals).

Since the introduction of the Taguchi method of industrial process design [38], modern businesses integrate process design with operations management. A unique feature of the present study is the incorporation of data driven nonlinear production routines within the SCN kernel with an efficient algorithm.

## 2. Literature Review

### 2.1 Uncertainties in SCN and their classification

Uncertainties in a SCN can be traced back to unexpected and erratic (impulsive) changes potentially affecting all connecting nodes of the 4-stage environmental-demand-economic-social operations. The nature of these perturbations is often unfamiliar and may or may not be time varying. In general, the reported literature shows uncertainty due to discrepancies in expected and actual role performance, material deficiencies, machine failures, defective product design, defects in operational system design, accidents at work place, shortcomings in demand forecasting, demand shocks, defects in the methods of inventory planning, inadequate investment planning, etc. [6, 26]. Based on the four pillars of supply chain sustainability, uncertainties can be taxonomically re-oriented as – a) Environmental uncertainties [26]; b) Demand uncertainties [20]; c) Social uncertainties [14] and d) Economic uncertainties [42].

### 2.1.1 Environmental Uncertainty

Analysis of environmental uncertainties in supply chain modeling is a well-researched topic [26]. The concept owes to Thompson and later coworkers [35,39General environmental uncertainty [37] can be associated with both negative and positive impacts [26]. Most of these studies focus on corporate environment rather than the internal grid of the supply chain network specifically.

### 2.1.2 Demand Uncertainty

Uncertainty is of two kinds: a deterministic uncertainty arising from lack of knowledge or misspecification of the models used in demand forecasting [6, 20], and stochastic uncertainty arising from unexpected external shocks such as demand uncertainty due to Covid pandemic [16]. Basically, optimization models with mixed integer programming and closed loop control networks are designed for supply-chain management. Demand uncertainty is then added as an additional feature to develop different scenarios for SCN management under demand uncertainty [20]. The COVID-19 pandemic has underlined the importance of demand uncertainty in SCN in a fluctuating market [19]. Our model incorporates demand uncertainty as one of the 4 structural pillars of the utility function.

*2.1.3 Social and Economic Uncertainty*

For an SCN that includes internal production operations economic and social uncertainties need to be modeled. These uncertainties arise from actions of multiple agents, dealing with maintenance of adequate credit worthiness, managing the inventories, revenue planning through price fixation, etc. and under demand uncertainty. [22]. Multi-objective nonlinear programming [22], general Lagrangian architecture to analyze build-to-order SCNs [44,45], have been designed to handle such demand uncertainties. In particular, the SCN constraints for build-to-order kernels have been enacted through Lagrangian multipliers, a more robust version of which has been adapted in this present study.

*2.1.4 Traditional Approaches towards Supply Chain Uncertainty Optimization*

Modeling the impact of uncertainties in a complex SCN is an active area of Supply chain management. Applications range from game theory [18], robust optimization [20], mixed integer linear programming [6], agent-based modeling [46] etc. Option pricing model using game theory is a preferred model for price discrimination under demand uncertainty for improving revenue management and increase supply chain efficiency [13]. Hu et al. [18] used a Stakelberg game structure to derive optimal ordering policy in a stochastic decentralized supply chain system. A similar approach is used to study the pricing and retail service decisions in a supply chain operating under fuzzy uncertainty environments to determine the equilibrium solutions both analytically and numerically to develop marketing strategies [43].

Agent-based modeling is another popular approach. Asim et al. [1] used fuzzy goal-programming to optimize a multi-echelon, multi-variable production-transportation closed loop SCN under real life uncertainty. Mekki et al. [46] used Belied-Desire-Intention (BDI) agent-based modeling considering three pillars of sustainability (environmental, economic, and social) for sustainability of SME supply chain under uncertainty. Shen [34] investigated the environmental impact of button batteries in closed-loop supply chain network under uncertainty. Shen proposed multi-objective mixed integer programming models, combining expected value model and chance-constrained model under uncertainty with a Life Cycle Assessment model to estimate the environmental effects on the SCN.

While successful in qualitative understanding of SCN networks, the traditional mathematical modeling techniques focusing primarily on strategies and decision making were not sufficiently probabilistically adapted to calibrate the performance of a SCN against parameter fluctuations and also with time. The present study will complement this knowledge gap by utilizing dynamic models that are of the form of nonlinear stochastic control problems that employ tools drawn from Lagrangian mechanics, including Lagrange multipliers that will be used to implement the subjective SME constraints.

*2.2 Existing Knowledge Gap and Main Contribution of this Study*

This article outlines a mathematical structure to minimize the risk incorporated in a four-dimensional cost function subject to dynamic production and operational structure of a SCN to assure sustainable production under uncertainty.

- Instead of subjectively adding extant parameters to incorporate uncertainty, our model incorporates uncertainty within the cost function itself.
- The study unifies all four types of uncertainties, with a focus on the four pillars of sustainability, in a single cost function. This is a major advancement to previous studies, e.g., Mota et al. [24] that relied on segmented analysis only.
- The model lays particular emphasis on understanding sustainability both as a deterministic construct as also a stochastic component, the latter when analyzed within a network.
- Our methodology employs an efficient algorithm that numerically analyzes the model to within a minute's resolution in performance (using our own Matlab codes). Starting from a general nonlinear model, we analyze the dynamics around the linearly stable points of the consequent Lagrangian equations of motion.
- The study precludes a major crossover from statistical analysis to predictive dynamical analysis where data from past market performance can predict the future sustainability of the SME.

**3. Methods**

*3.1 Problem Statement*

Uncertainties in its supply chain can drive an SME towards bankruptcy. The actual survival of Small and Medium sized Enterprises (SMEs) operating on shoe-string budgets is a major technical challenge. Since even a minor fluctuation in the markets could cause a major flutter in the fortune of an SME, the real benchmark of a quantitative SCN is its ability to match predictions against real numbers. As we show, this can only be achieved from a precise understanding of the (four) phenomenological sources of uncertainty and their cumulative effects from a holistic model. The goal of this paper is to develop this mathematical model incorporating all four pillars (EDES) of uncertainties, then analyze and rank their interdependencies. The emergent toolbox will grade the sources of uncertainty, quantify their contributions in the market, and identify safety windows that ensure the safe operations principle of the SME concerned.

*3.2 Model Description*

The model is inspired by the four uncertainty categories pertaining to the SCN of any company which are identified via literature review, brainstorming and case studies [20]. In this study, we consider a generalized version of a supply chain starting from the raw material procurement and ending with the production line delivery to the market, translating the mechanics to a cradle-to-utilization model (Figure 1). Our minimalist cost function-based model utilizes the central structure discussed in the following Section 3.2.2 (Eq. 10) where contributions from individual sectors are weighted against their respective uncertainty. This ensures the robustness of the prediction.

*3.2.1 Assumptions*

The model relies on some practical, often common sense, inputs:

i) For daily statistics, yearly data is divided by 3000 (INR), that assumes 10 hours' work for each working day, over 300 working days in a year.

ii) Legislation and miscellaneous costs are considered unchanged, but the annual cost is taken (allocated to each day by dividing the annual cost by 3000 INR).

iii) Unit costs have been assumed constants per case, as these remained relatively unchanged during the duration of this study.

*3.2.2 The Cost Function Model*

The model presented in this paper targets a cost minimization model with four horizontal cost components associated with environmental, economic, social, and demand factors with associated uncertainties, arranged into a hierarchical pattern determined by Saaty's Analytical Hierarchical Programming. The model uses "cost function" as a "free energy" as it captures the randomness induced by various components of the SCN." As economists' normally postulate a U-shaped convex average cost function, we assume quadratic convex cost function to ensure linear relationships for cost minimizing stable solutions [25].

*Cost components*

F = Cost Function

$C_{Environment}$ = Cost function component for Environmental Uncertainty

$C_{Social}$ = Cost function component for Social Uncertainty

$C_{Economic}$ = Cost function component for Economic Uncertainty

$C_{Demand}$ = Cost function component for Demand Uncertainty

The variables we use in the model are:

$V_{CO2}$ = Amount of $CO_2$ generated

$H_p$ = Energy Consumption due to processes involved

$W_p$ = Water used due to the processes involved

$W_w$ = Wastewater produced in the whole process

$N_1$ = No. of labors

$N_2$ = No. of employees

$N_3$ = No. of social responsibility activities in a year

$N_4$ = No. of products sold

$N_5$ = No. of operations involved

$N_6$ = No. of type of taxes

$N_7$ = No. of shipments

$N_8$ = No. of other logistics

For deriving the interdependencies amongst these variables, Saaty's Analytical Hierarchical Process (AHP) is used [4, 32]. The Analytic Hierarchy Process (AHP) is a method of developing a common measure combining discrete and continuous variables through pairwise comparisons. It has found its widest applications in multicriteria decision making, planning and resource allocation and in conflict resolution. In (Online) Appendix 1, we present details about AHP and the algorithm used. The working algorithm follows that of Saaty [32] as detailed in (Online) Appendix 2. Two AHPs are used: the first (AHP1) prioritizes the uncertainty variables, followed by a second (AHP2) that is a layered AHP (with two layers of alternatives) that prioritizes the interdependence between variables. The normalized eigenvalues obtained from the AHP are taken as the weight factors. These eigenvalues have been rated from the data obtained from an Indian SME (anonymous) which is primarily a manufacturing SME in West Bengal, India. Usually, whatever method is used to derive these weights, they depend on the algorithm used and hence are subject to standard errors. Perturbation analysis is performed (Using Equations (8) and (11) below)) to ensure a confidence level ≥ 90%.

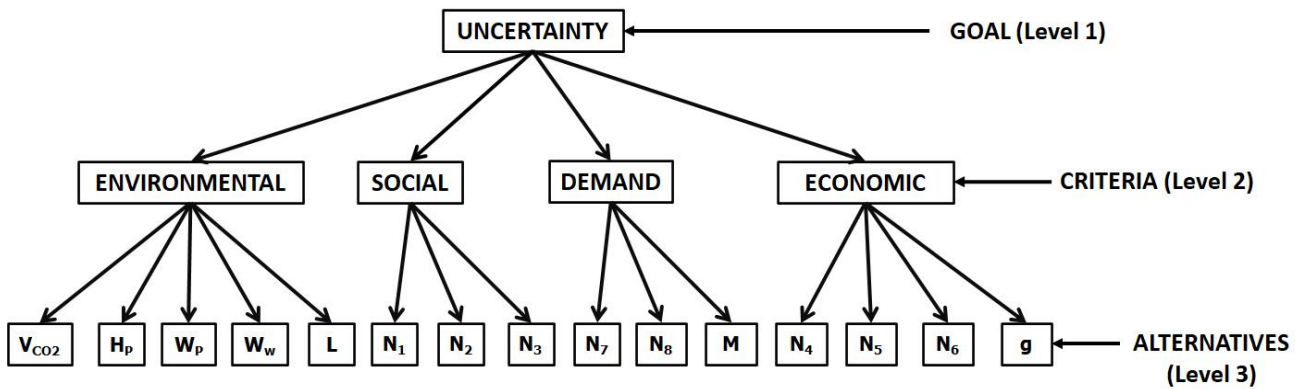

**Figure 3**: AHP model 1 to determine the general alterative rankings. Volume of $CO_2$ generated ($V_{CO2}$), Energy consumption due to processes involved ($H_P$), Water used due to the processes involved ($W_P$), Wastewater produced in the whole process ($W_W$), Legislative Costs (L) are the alternatives connected to Environmental node; No. of labors ($N_1$), No. of employees ($N_2$), Number of social responsibility activities in a year ($N_3$) are the alternatives connected to Social node; No. of transportations involved (N7), No. of other logistics (N8), Miscellaneous cost (M) are the alternatives connected to the Demand node; No. of products sold (N4), No. of operations involved (N5), No. of type of taxes (N6), Disaster management fund (g) are the alternatives connected to economic node.

AHP1 is a hierarchical structure that provides the alternative rankings (i.e., the concerned variables for the uncertainties); this has been used as weights/coefficients ($A_1$ to $A_{15}$) of the objective function F. It consists of a goal cluster, a criteria cluster and an alternatives cluster (Figure 3). AHP2 is a layered diamond like hierarchical structure involving a goal cluster, a criteria cluster and two layered alternatives (Figure 4). The layered structure connects the interdependent parameters and variables through a meshed network (details in Online Appendix 2).

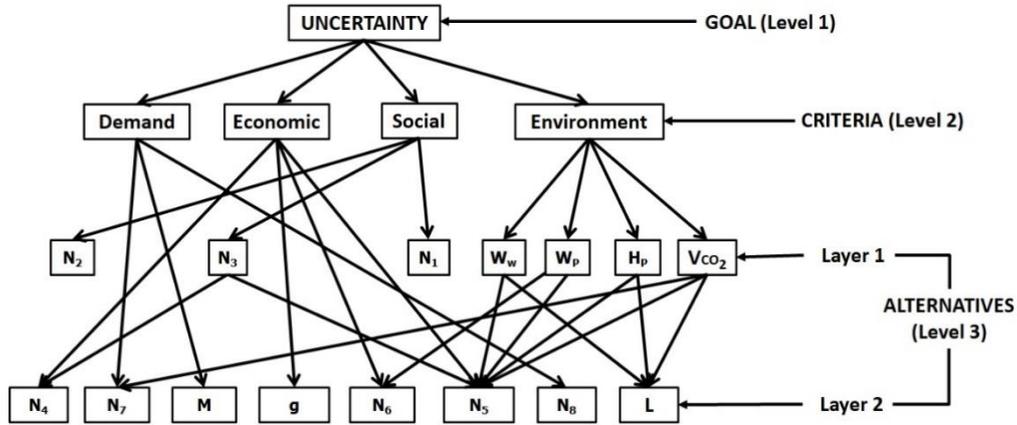

**Figure 4**: Layered AHP model for determination of interrelationship values

*3.2.3 Inter-dependency of the variables*

The interdependencies between these variables, based on AHP, are described by the following equations:

$$V_{CO_2} = a_1 L + a_2 N_5 + a_3 N_7 + a_{12} L N_5 + a_{23} N_5 N_7 + a_{31} L N_7 + a_1' L^2 + a_2' N_5^2 + a_3' N_7^2$$
$$W_P = W_P^O + b_1 N_5 + b_2 N_5^2$$
$$H_P = c_1 L + c_2 N_5 + c_{12} L N_5 + c_1' L^2 + c_2' N_5^2$$
$$W_w = d_1 L + d_2 N_5 + d_{12} L N_5 + d_1' L^2 + d_2' N_5^2$$
$$N_3 = \alpha_1 N_4 + \alpha_2 N_6 + \alpha_{12} N_4 N_6 + \alpha_1' N_4^2 + \alpha_2' N_5^2$$
$$N_4 = \beta_1 N_7 + \beta_2 N_8 + \beta_{12} N_7 N_8 + \beta_1' N_7^2 + \beta_2' N_8^2$$
$$N_7 = \gamma V_{CO_2}$$

In arriving at the above formulation, we have restricted upto quadratic relationship, including cross-coupling between variables. This ensures incorporation of nonlinear correction at least at the minimal level.

*Parameters used in the model are:*

$f_1$ = Cost for $CO_2$ recovery

$f_2$ = Cost for thermal pollution prevention

$f_3$ = Cost for water used

$f_4$ = Wage of one labour

$f_5$ = Salary of one employee

$f_6$ = Average unit cost for social responsibility activity

$f_7$ = Revenue earned from product sold

$f_8$ = Cost of each operation

$f_9$ = Average cost for disaster management per day

$f_{10}$ = Cost for transportation

$f_{11}$ = Cost for other logistics

y = Cost for wastewater treatment

M = Miscellaneous cost

L = Legislative Costs

T = Cost for Taxes

g = Disaster management fund

*Weight factors (parameters)*

$\epsilon_i$'s = Weight factor for the four cost functions derived from AHP (where i = 1 to 4)

$A_i$'s = Weight factor for the main parameters derived from AHP (where i = 1 to 15)

$a_i$'s, $a_{ij}$'s & $a'_i$s = Interdependency values derived from AHP for $V_{CO2}$ (where i = 1 to 3 & j = 1 to 3)

$b_i$'s = Interdependency values derived from AHP for $W_P$ (where i= 1 to 2)

$c_i$'s, $c_{ij}$'s & $c'_i$s = Interdependency values derived from AHP for $H_P$ (where i= 1 to 2)

$d_i$'s, $d_{ij}$'s & $d'_i$s = Interdependency values derived from AHP for $W_w$ (where i= 1 to 2)

$\alpha_i$'s, $\alpha_{ij}$'s & $\alpha'_i$s = Interdependency values derived from AHP for $N_3$ (where i= 1 to 2)

$\beta_i$'s, $\beta_{ij}$'s & $\beta'_i$s = Interdependency values derived from AHP for $N_4$ (where i= 1 to 2)

$\gamma$ = Interdependency value derived from AHP for $N_7$

As can be seen the variables span all four dimensions of supply chain sustainability and are critically chosen to ensure tracking of all modes of perturbations affecting the supply chain, ranging from environmental pollution aspects to socio-economic features primed for generating strategies for sustainable production. Details on the data sources are available in the Supplementary document (Table S5) in (Online) Appendix 1.

$$F = C_{Environment} + C_{Social} + C_{Economic} + C_{Demand}, \text{ where} \qquad (1)$$

Going by its origin, the cost function is a product of two variables, typically the volume of a certain quantity generated (e.g. $V_{CO2}$) times its unit production cost (e.g. $f_1$). This naturally requires two running indices of summation. However, with each running index varying between 1 to 300, e.g. 1 < i < 300, and with 15 variables in operation, this would lead to an unmanageable situation mostly due to lack of information concerning the off-diagonal terms in the product matrices. We thus opted for an approximate but minimalist model using inputs from the SME leaders and subject experts focusing on the key diagonal entries only. Technically, this amounts to the reduction of a double index module to one of single index. In the following equations the summation runs over all the 300 working days of the plant, while the variables are taken as daily values of those variables.

$$C_{Environment} = \sum_{i=1}^{300} V_{CO_{2_i}} f_{1_i} + \sum_{i=1}^{300} H_{P_i} f_{2_i} + \sum_{i=1}^{300} W_{P_i} f_{3_i} + \sum_{i=1}^{300} W_{w_i} y_i + \sum_{i=1}^{300} L_i, \qquad (2)$$

where $f_1$ is the cost for carbon dioxide mitigation/recovery, $f_2$ is the cost for energy recovery/minimization, $f_3$ is the cost for water consumption, y is the cost of wastewater treatment and 1<i<300, referring to the number of yearly working days (serving as the upper limit of the summations).

$$C_{Social} = \sum_{i=1}^{300} N_{1_i} f_{4_i} + \sum_{i=1}^{300} N_{2_i} f_{5_i} + \sum_{i=1}^{300} N_{3_i} f_{6_i}, \qquad (3)$$

where $f_4$ is the wage of one labor, $f_5$ is the salary of one employee, $f_6$ is the average cost for social responsibility activity.

$$C_{Economic} = \sum_{i=1}^{300} N_{4_i} f_{7_i} - \sum_{i=1}^{300} N_{5_i} f_{8_i} - \sum_{i=1}^{300} g_i f_{9_i} - \sum_{i=1}^{300} N_{6_i} T_i, \qquad (4)$$

where $f_7$ is the revenue earned from the product sold, $f_8$ is the cost for unit operations, $f_9$ is the average cost for disaster management per day.

$$C_{Demand} = \sum_{i=1}^{300} N_{7_i} f_{10_i} + \sum_{i=1}^{300} N_{8_i} f_{11_i} + \sum_{i=1}^{300} M_i, \qquad (5)$$

where $f_{10}$ is the cost of transportation, $f_{11}$ is the cost of other or special logistics, M is the miscellaneous cost.

Together with the relevant Analytical Hierarchy Process (AHP) weight factors, as detailed above, the cost function, serving as the free energy, can be structured as follows:

$$F = \epsilon_1 \left( \sum_{i=1}^{300} A_1 V_{CO_{2_i}} f_{1_i} + \sum_{i=1}^{300} A_2 H_{P_i} f_{2_i} + \sum_{i=1}^{300} A_3 W_{P_i} f_{3_i} + \sum_{i=1}^{300} A_4 W_{w_i} y_i + \sum_{i=1}^{300} A_5 L_i \right) +$$
$$\epsilon_2 \left( \sum_{i=1}^{300} A_6 N_{1_i} f_{4_i} + \sum_{i=1}^{300} A_7 N_{2_i} f_{5_i} + \sum_{i=1}^{300} A_8 N_{3_i} f_{6_i} \right) + \epsilon_3 \left( \sum_{i=1}^{300} A_9 N_{4_i} f_{7_i} - \sum_{i=1}^{300} A_{10} N_{5_i} f_{8_i} - \right.$$
$$\left. \sum_{i=1}^{300} A_{11} g_i f_{9_i} - \sum_{i=1}^{300} A_{12} N_{6_i} T_i \right) + \epsilon_4 \left( \sum_{i=1}^{300} A_{13} N_{7_i} f_{10_i} + \sum_{i=1}^{300} A_{14} N_{8_i} f_{11_i} + \sum_{i=1}^{300} A_{15} M_i \right). \qquad (6)$$

Here $\epsilon_1, \epsilon_2, \epsilon_3, \epsilon_4$ are the weights corresponding to individual uncertainties and $A_i$'s are the weights of the individual variables. In absence of real market data, these too have been obtained through AHP analysis (details in Online Appendix 1). This is the overall generic cost function (eq. 6). In order to reduce the energy (often related to the optimized entropy of a chaotic system) of the system (i.e., the supply chain network uncertainties represented by the mathematical model), the dominant dynamic variables are only considered at the final stage (section 3.3.1) based on subjective (industrial partners) inputs. To understand the dynamics of the system, Lagrange Multipliers are introduced to develop a constrained system (section 3.3.2) in line with [44, 45]. In both cases, the cost function takes a modified form, which are detailed in the latter sections. To avoid any confusion, further details are attached in (Online) Appendix 2 (following on from Online Appendix 1).

*3.3. Optimization Algorithm*

In order to arrive at an optimal strategy for the SCN, the model cost function (Eq. 6) is optimized (minimized). The resultant Hessian matrix, like an information matrix in statistics, characterizes the general uncertainty scenario of a supply chain network. The optimization mechanism runs through two gateways – (a) Unconstrained Optimization, dealing largely with a hypothetical scenario in which the SCN has unlimited resources in finances, workforce, machinery and equipment and raw material and (b) Constrained Optimization, which portrays a more realistic scenario of a company with limited resources. To arrive at a dynamical formulation of the SCN, we optimize the cost function with a time independent Lagrangian structure (using textbook treatment of classical mechanics as in [15, 17]). We assume the Lagrangian to be time independent as it is our assumption that we are dealing with a SCN associated with a SME that is in a stabilized working condition, after going through its teething problems, and not yet ready to grow into a bigger company through a dynamic growth path. Equations (13) and (16) are solved using data obtained from an anonymous Indian Small and Medium Enterprise (SME) company. The following algorithmic flowchart outlines the problem to solution path.

**STEP 1:** Identify factors imposing uncertainty along the supply chain network & Real-life information.

**STEP 2:** Develop the Cost Function Model and convert into an optimization problem (Eq. 6 and Online Appendix 2).

**STEP 3:** Derive the weight factors & interrelation coefficients using AHP (Section 3.4.1 & Online Appendix 1).

**STEP 4:** Evaluate the first and second derivatives and build the hessian matrix (Online Appendix 2).

**STEP 5:** Evaluate the determinant value of the Hessian H (Online Appendix 2).

**STEP 6:** Check for det(H)>0 (for a cost minimization problem). If yes, then go to step 7, else validate data and model, go to step 1.

**STEP 7:** Identify the effective uncertainty variables (Online Appendix 2).

**STEP 8:** Reduce the uncertainty variables subjected to constraints (Eq. 8).

**STEP 9:** Use the Euler-Lagrange formulation to define dynamical model (Eq. 9).

**STEP 10:** Introduce the Lagrange Multipliers to analyze the constrained problem (Eq. 11).

**STEP 11:** Derive the elements of the matrix using real data (Online Appendix 2).

**STEP 12:** Identify the boundary conditions (Table 1).

**STEP 13:** Solve constrained & unconstrained problems simultaneously.

Independent verification is attained using MATLAB R2021a (bvp4c) and Mathematica 12 (structured Runge-Kutta 4) to solve the corresponding boundary value problems.

*3.3.1 Unconstrained Problem*

We follow the standard method to generate the dynamic path from the steady state equilibrium. We use the free energy F (The total cost of uncertainty) and employ Lagrangian mechanics (details in Online Appendix 2). The time dependent movements of the cost function modules are expected to satisfy optimum cost function formulation to generate a steady state solution. A perturbation of the SCN will 'kick' the system out of its steady state equilibrium. Then the system will either follow a conservative trajectory towards that steady state (a stable fixed point) or follow a non-conserved dynamics towards an unstable fixed point. To mathematically encapsulate this, we prescribe the multivariate Euler-Lagrange structure (Step 9) ([15], [28]) $\delta\left(\frac{\partial F}{\partial N_i}\right) = \delta\left(\frac{d}{dt}\frac{\partial F}{\partial \dot{N}_i}\right)$, where F is the Free Energy or cost in our example. Then, in terms of the leading dynamical variables (identified in consultation with the SME) $(N_4, N_5, N_7, VCO_2)$, the cost function kinetics runs as a 'tug of war' between production, cost and delivery, each changing with time (details in Online Appendix 2). Focusing on the 4 dominant dynamical variables $(N_4, N_5, N_7, VCO_2)$, Euler-optimization of Eq. (6) leads to

$$\delta\left(\frac{d}{dt}\begin{bmatrix}\frac{\partial F}{\partial N_4}\\ \frac{\partial F}{\partial N_5}\\ \frac{\partial F}{\partial N_7}\\ \frac{\partial F}{\partial V_{CO_2}}\end{bmatrix}\right) = \begin{bmatrix} m_{11} & 0 & m_{13} & 0 \\ 0 & m_{22} & m_{23} & 0 \\ 0 & m_{32} & m_{33} & 0 \\ 0 & m_{42} & 0 & m_{44} \end{bmatrix}\begin{bmatrix}\delta N_4\\ \delta N_5\\ \delta N_7\\ \delta V_{CO_2}\end{bmatrix}$$

(7)

which translates to the following dynamical model

$$\frac{d^2}{dt^2}\begin{bmatrix}\xi_1 \delta N_4\\ \xi_2 \delta N_5\\ \xi_3 \delta N_7\\ \xi_4 \delta V_{CO_2}\end{bmatrix} = \begin{bmatrix} m_{11} & 0 & m_{13} & 0 \\ 0 & m_{22} & m_{23} & 0 \\ 0 & m_{32} & m_{33} & 0 \\ 0 & m_{42} & 0 & m_{44} \end{bmatrix}\begin{bmatrix}\delta N_4\\ \delta N_5\\ \delta N_7\\ \delta V_{CO_2}\end{bmatrix},$$

(8)

Here $\delta N_4, \delta N_5, etc.$ represent perturbations of variables $N_4, N_5$, etc. around their equilibrium points, where 'equilibrium points' refer to the 'break even trade situation'. Eq. (8) is reminiscent of Newton's law of motion that describe how fast a system can accelerate or decelerate away from its equilibrium, that 'break even' point. In other words, a trajectory away from the equilibrium manifold is indicative of possible bankruptcy and vice versa.

*3.3.2. Constrained Problem*

A real supply chain problem is a constrained model that compares its cost function against the limitations imposed by the constraints like budget constraint, restrictions on the number & quality of work forces, transport and others. To analyze such constrained problems, a Lagrangian structure for the cost function has been used where constraints were introduced through the established Lagrange multiplier formalism ([30]), allowing for parameterization in terms of the constraints ([15]). Our Lagrangian 'L' is defined as follows:

$$L = F - \lambda_1(N_1 f_4 + N_2 f_5 - C) - \lambda_2(N_4 f_7 - E) - \lambda_3(V_{CO_2} f_1 - V) - \lambda_4(N_3 f_6 - R), \qquad (9)$$

where the Lagrange multipliers $\lambda_i$'s represent the additional utility cost ('friction') arising due to constraints on the 4 variables $N_4$, $N_5$, $N_7$, $VCO_2$. The quantity C represents the maximum budget allowed for the wages; E is the maximum expected earning based on products sold; V is the carbon footprint cost; R identifies the maximum allowable cost for social activities (which is partially rebated as tax benefit). Note that the original cost function F refers to 15 variables (n=15 as in Online Appendix 1, Step 2) and 3 levels (as in Figures 3 and 4). Together, they define the dynamical evolution of the green supply chain as a mechanistic model:

$$\delta\left(\frac{d}{dt}\begin{bmatrix} \frac{\partial L}{\partial N_4} \\ \frac{\partial L}{\partial N_5} \\ \frac{\partial L}{\partial N_7} \\ \frac{\partial L}{\partial V_{CO_2}} \end{bmatrix}\right) = \begin{bmatrix} k_{11} & 0 & k_{13} & 0 \\ 0 & k_{22} & k_{23} & 0 \\ 0 & k_{32} & k_{33} & 0 \\ 0 & k_{42} & 0 & k_{44} \end{bmatrix} \begin{bmatrix} \delta N_4 \\ \delta N_5 \\ \delta N_7 \\ \delta V_{CO_2} \end{bmatrix} \qquad (10)$$

As before, in terms of the real variables, the cost optimized SCN kinematics abides the following equation

$$\frac{d^2}{dt^2}\begin{bmatrix} \psi_1 \delta N_4 \\ \psi_2 \delta N_5 \\ \psi_3 \delta N_7 \\ \psi_4 \delta V_{CO_2} \end{bmatrix} = \begin{bmatrix} k_{11} & 0 & k_{13} & 0 \\ 0 & k_{22} & k_{23} & 0 \\ 0 & k_{32} & k_{33} & 0 \\ 0 & k_{42} & 0 & k_{44} \end{bmatrix} \begin{bmatrix} \delta N_4 \\ \delta N_5 \\ \delta N_7 \\ \delta V_{CO_2} \end{bmatrix} \qquad (11)$$

## 4. Results and Discussions

In this section, the time variation of the key parameters has been discussed, respectively for the constrained (dystopian) and unconstrained systems (utopian). The main variables sensitive to this system are the number of products sold ($N_4$) and volume of carbon dioxide emission ($VCO_2$). These two variables are the indicators of the sustainable production line which we use to identify the "operation-windows" for the SMEs policy implementation and decision making. Here, we analyze the carbon footprint of an SME as a function of varying cost functions to identify the best optimized strategy for an SME to cope with such environmental demands against their stringent budgets. In terms of our model, the effects of the correlation (specific to a certain case that may vary with change in companies) of $N_4$ and $VCO_2$ with business growth will reflect the best $CO_2$ emission minimization-versus-profit optimization scenario. The trade-off between these two rival variables defines the optimization strategy by identifying an optimal time point when the SME can flourish while also ensuring environmental sustainability.

This study considers two different strategies that a SME can opt for, mathematically represented as two different boundary conditions (Table 2). The successive subsections present two case studies. These case studies demonstrate the effect of model-based decision making on performance (sustained business or bankruptcy). We adopt a minimalist approach where maximally varying variables/parameters are identified and then analyzed. All constrained parameters are identified through extra 'c's alongside the main variable, e.g., $Nc_4$ i.e., product sold in a constrained environment. In reality, several occasions may arise where the SMEs have to identify subjective strategies to control their business, again mathematically represented by appropriate Initial Conditions (IC) and Boundary Conditions (BC).

**Table 1: Boundary Conditions for Evaluation of Results**

| Sl. | $N_4$ | $N_5$ | $N_7$ | $VCO_2$ | $VCO_2$ Check | |
|---|---|---|---|---|---|---|
| | | | | | 3-Y | 5-Y |
| | Product Sold (% of sales) | Operations Involved (No.) | Transportation modes (No.) | Volume of $CO_2$ (tons of $CO_2$ Eq.) | (tons of $CO_2$ Eq.) | (tons of $CO_2$ Eq.) |
| | *IC (Initial Conditions)* | | | | | |
| 1 | 0.3 | 76 | 2 | 2.86 | 3 | 2.3 |
| 2 | 0.3 | 76 | 2 | 2.86 | 2.19 | 1.46 |
| | *BC (Boundary Conditions)* | | | | | |
| 1 | 0.5 | 80 | 5 | 5.2 | 2.4 | 2 |
| 2 | 0.5 | 82 | $\frac{dN_7}{dt} = 0$ | 2 | 1.72 | 1 |

In case of $N_4$ i.e., product sales, we have considered percentage of product sales as ICs and BCs. Here, 0.3 and 0.5 as IC and BC respectively indicate 30% of the product sales as the initial point and 50% as the terminal point. $N_5$ and $N_7$ are obtained from the SME data. For $VCO_2$, the unit used is tons of $CO_2$ equivalent.

**4.1** *Case Study 1: The SME chooses to increase the product sale without compromising on the carbon emission reduction (i.e., carbon emission increases)*

This is a classic case of non-compliance where the SME chooses to increase the production and ignore emission aiming to flourish by increased sales. However, they are economically constrained, not to be able to invest to mitigate increasing $CO_2$ emission. As a result, the boundary conditions change not only for these two variables but also others such as transportation cost increases. This case complements SDG 9, 13-15 (https://sdgs.un.org/goals) i.e., industry, innovation and infrastructure; climate action, life on land and life under water.

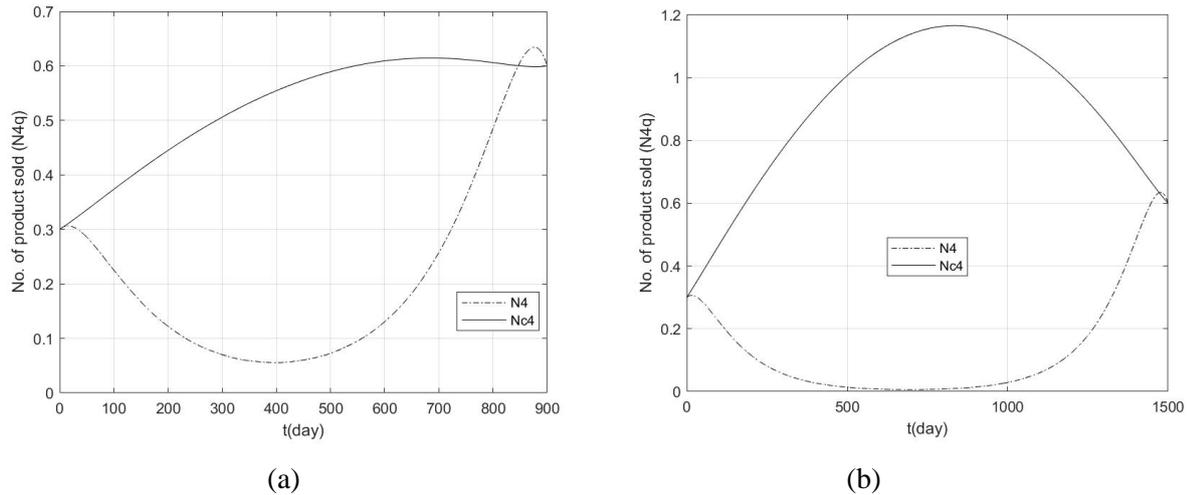

(a)                                                         (b)

**Figure 5**: **Time dependence of $N_4$ in unconstrained (dotted line) and constrained (solid line) environment, plots obtained from simultaneous solution of Eqs. (8) and (11): (a) 3-year time span; (b) 5-year time span when SME ignores emission.**

In the unconstrained environment, the company starts to lose money and reaches its minimum in the first quarter of the second year (Figures 5a). Eventually, it gains momentum by raising their sale and exceeds the target slightly and then reaches the goal by end of the third year. However, in the 5-year time span (Figures 5b), the curve flattens and then rises after the third year indicating that even in the most favorable of conditions, the strategy will not give the SME best results in the long run. The realistic version of the results shows that in a 3-year time frame, the company smoothly matches its projected target within the stipulated timeline. In the longer run (5-year time frame), our model predicts that this company will start depreciating and will only break even after the second year. However, it rises to its pinnacle towards the end of the third year as it exceeds the target. Thereafter, the curve steeply falls to reach its target value.

In an ideal world allowing for investment over an infinitely large time period (10+ years), this strategy will fail to give good results as the curve will become oscillatory and hence unstable. The results clearly indicate that the rise and fall in sales are due to demand uncertainty and the SME needs to take further actions to flourish. This is a predictive perspective that we arrive at from our model analysis. An inbuilt assumption to all such analyzes is the zero-uncertainty enforcement at the end points of a boundary value system, a situation that mathematically mimics a certainty in a decision before a process starts and after it ends, a logical paradigm. This is, admittedly, an over-optimistic approach to encourage product sales, but this strategy fails. The change in strategy is the crucial step here that can help the SME to slowly move towards a sustainable production regime.

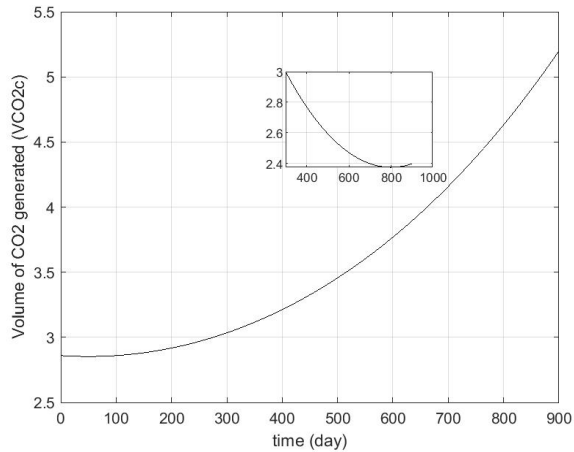

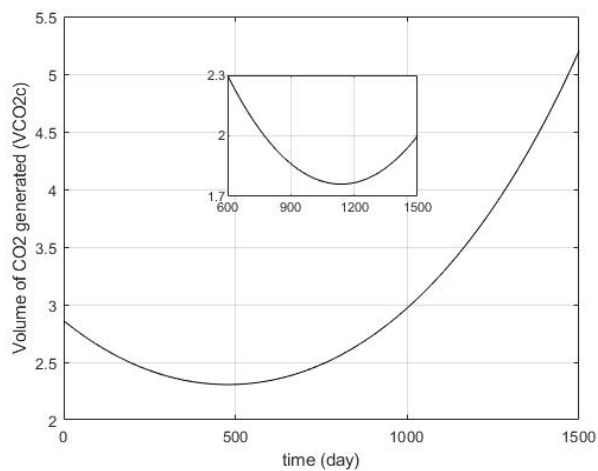

(a)                                                                                   (b)

**Figure 6: Time Dependence of $V_{CO_2}$ in a constrained environment, plots obtained from simultaneous solution of Eqs. (8) and (11): (a) 3-year time span & (b) 5-year time span, when SME ignores emission**

In the unconstrained environment, $VCO_2$ gives non-converging solutions which imply that without carbon emission control, the system is unstable, and hence unsustainable. The sample curve for the unconstrained environment is provided in (Online) Appendix 2. As expected, the emission curve rises to its peak value in a parabolic profile at both 3 (Figures 6a) and 5-year (Figures 6b) timelines, implying that the adopted strategy is not environmentally sustainable. To address this, our model advises the company to re-strategize the carbon footprint at the end of each relevant time period, identified by the minimum in the respective plots. The results are shown in the insets of figure 6(a) and 6(b). In the 3 years' timeframe, when the SME actively targets a reduction in its carbon emission by the end of the first year (strategizing in advance), an overall reduction of ca 20% is possible by the end of the third year. Similarly, in the 5 years' time frame, the timeline for a similar action will be the end of the 2-year timeline that then contributes to ca 22% reduction of $CO_2$ emission by the end of the fourth year. Such recursive monitoring of strategy could continue to improve the carbon footprint in successive years, highlighting a major achievement of this optimal model. *This case study suggests that our model not only complements SDG 13, i.e., climate action, but also addresses SDG 15 which emphasizes the*

*impact of life on land. As an SME makes effort to reduce the CO$_2$ emissions to meet a preset target, it is essentially contributing towards negative climate change inspiring better life sustenance.*

**4.2 *Case Study 2: The SME chooses to increase the product sale without increasing the logistics cost with a certain compromise on the carbon reduction (i.e. carbon emission decreases)***

This is a unique case where the SME chooses to increase the production and aims to flourish by bringing up the sale and focuses on carbon reduction as well. However, they are not willing to further add up on the logistics cost. Mathematically, this changes the boundary conditions for the associated two variables, together with an increase in the number of operations. In this case, the output variables are associated with SDG 9, 11, 12 and 13.

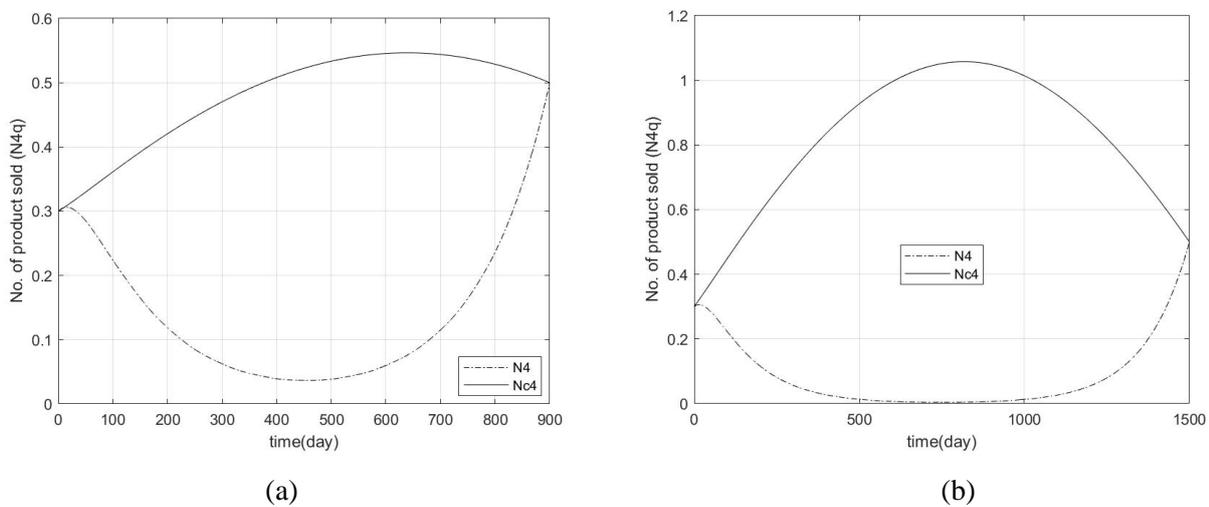

(a)            (b)

**Figure 7: Time dependence of N$_4$ in unconstrained (dotted line) and constrained (solid line) environment, plots obtained from simultaneous solution of Eqs. (8) and (11): (a) 3-year time span; (b) 5-year time, when SME focuses on emission reduction**

In the unconstrained environment, the product sale decreases to a minimum of less than 10% during the middle of the second year and eventually recovers to reach the goal in the end of the third year (Figure 7a). Compared to its 3-year strategy, the 5-year strategy shows higher time period of loss which is during 500 < time < 900 and then exponential recovery to reach goal (Figure 7b). This suggests that, in arbitrage condition, the strategy might reach the goal, but the SME may be operating in a break-even state for a significant amount of time. This again demands a change in policy for the SME at a certain interval to reinvigorate product sales.

In the constrained system, the product sale of the SME increases exponentially and reaches a maximum of 55% at the end of the second year and reaches the goal in the end of the time period of three years. Conversely, in the 5-year strategy, product sale increases and exceeds 100% in the end of the third year. The sale decreases thereafter and reaches the prefixed boundary values. As long as it does not affect the production and economic sustainability of the SME, the strategy may prevail. Our model clearly identifies the action points for the SME. The maxima of the curves represent time points when the SME must intervene with an appropriate change in policy. It is suggested that regular monitoring and cross-validation of existing policy should be carried out for business sustainability. As identified by the model, the action points are the most important takeaway, as those represents the point of inflection. Hence, the model handovers the control regime to the supply chain managers/the SME to take necessary action for achieving a sustainable production line.

It is imperative from the results obtained that increasing product sales will not only increase resource consumption (SDG 12) but will also impact on climate change (SDG 13) due to carbon emissions from the processes involved. This will negatively impact life on land and water (SDG 14 and 15). Our work focuses on the optimization of logistics cost that will partly offset this negative input bearing in mind that technologically such emissions are not completely avoidable. As suggested, policy changes and regular monitoring will certainly lead to positive effects towards achieving the SDGs. Sustainable consumption of resources will not only contribute towards positive impacts on SDG 12 – 15; but it will also address certain aspects of sustainable cities (SDG 11). Overall, there are enough scopes for partnership among goals (SDG 17) that can lead to decent economic growth (SDG 8).

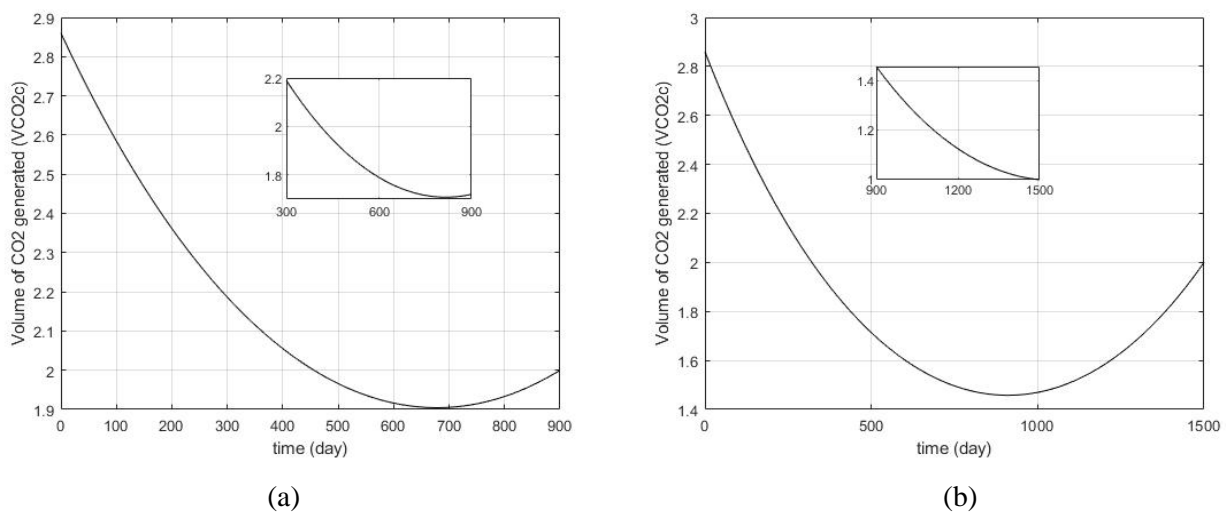

(a)           (b)

**Figure 8**: **Time Dependence of $VCO_2$ in a constrained environment, plots obtained from simultaneous solution of Eqs. (8) and (11): (a) 3-year time span & (b) 5-year time span, when SME focuses on emission reduction.**

In this case, the SME takes active initiatives to reduce carbon emission. The emission curve has a parabolic profile and attains a minimum before reaching its desired value both in the 3-year (Figure 8a) and 5-year (Figure 8b) time spans. Our model suggests that the SME can strategically reduce their carbon footprint by periodically reorienting the strategy after regular time intervals. However, in this case, the model does not specifically constrain a strategic revision at a specific cost (function)-minimum, rather the choice will remain with the SME to periodically monitor and strategize in advance. The results are provided in the insets of figure 8 (a) and 8 (b). In the 3-year timeframe, when the SME revises its carbon emission at the end of the first year, a reduction of ca 22% is achievable by the end of the third year. Likewise, in the 5-year timeframe, the target timeline for revision will be the end of the third year which makes it possible to reduce carbon emissions to ca 30% by the end of the fifth year. The case study also refers to a situation where both demand uncertainty and environmental uncertainty can be eliminated in a single intervention. Additionally, the general negligence of the SMEs towards carbon footprint reduction can be repaired without compromising the profit margin, as this specific case demonstrates. Similarly, cleaner production strategies can be useful in achieving a SME with sustainable consumption and production line. Hence, this case offers flexibility to jointly address SDG 8 and 13, and also SDG 9. The positive impact of this case will lead to sustainable cities (SDG 11) and life on land and under water (SDG 14 and 15).

**4.3 Discussions in line with Sustainable Production and SDGs**

The current investigation, though highly focused on supply chain analytics using the uncertainty-manifold as a metric (vector), the construct is a generalized structure that any industry can adopt for analyzing its sustainability performance and hence move towards a 'sustainability practice' approach. We use an SME as a case study. However, the model can still be easily adjusted with a different SME domain, for example, a chemical industry or a waste recycling plant, expectedly involving the dominant variables for that system. The findings of the study largely complement the Sustainable Development Goals (SDGs) 2030 adopted by the UN member countries in 2015. Based on the findings, we have identified eight SDGs which are being addressed by our study (as identified in un.org).

The two case studies provided in the previous subsections outlines interesting results. The major variables i.e., number of products sold ($N_4$) and volume of carbon dioxide emission ($VCO_2$) are key indicators to emphasize. In Case Study 1, the SME fails to achieve projected target even in favorable conditions. The sustainable production strategy for the SME is to choose its strategies based on market fluctuations and thereby reducing the risk of bankruptcy. From the environmental perspective, an SME may benefit from up to 22% reduction in the carbon footprint for which recursive monitoring is the key to success. SP strategies such as this one will not only enhance the environmental image but also ensure a clean and green SME.

On the other hand, Case Study 2 identifies the action point where SP strategies needs to be implemented. In this case, SP strategies may be represented by technological innovation or policy change. The environmental achievement is substantial and suggests that periodic threshold revisions will provide better results. Additionally, our model suggests that strong policy intervention will reduce carbon emissions. This will directly address SDG 13 and indirectly SDG 14 & 15.

In summary, our model can identify the action points for implementation of SP strategies as well as the SDGs it targets to address. Obviously, the numbers and evolved variables are subjective of the datasets of the current

SME, but the approach is generic enough for the technique to be ported to other challenging cases such as the waste management sector as well as biorefinery sectors.

## 5. Conclusions

Under the current investigation, we consider a generic "cradle-to-utilization" supply chain network. We develop a cost function model which considers four groups of uncertainties - Environmental, Economic, Social and Demand, identified from the four pillars of supply chain sustainability. The model also quantifies the interdependence of the uncertainty factors. Analytic Hierarchical Process (AHP) is employed to determine the weight factors and the analytical interdependencies. Euler-Lagrange formulation, an established framework in classical mechanics that mathematically outlines the time evolution of a dynamically evolving variable, is then used to define the time dynamics of the market and its interacting factors. Two cases are developed – a) Unconstrained optimization and b) constrained optimization. Both are studied within the same Euler-Lagrange framework. Further optimization is carried out by solving the boundary value problems using MATLAB and Mathematica.

The inputs and outputs of the model are all based on the 4 uncertainty pillars. Specifically with reference to the free energy F, the inputs are the variables Volume of $CO_2$ generated ($VCO_2$), Energy consumption due to processes involved ($H_P$), Water used due to the processes involved ($W_P$), Wastewater produced ($W_W$), No. of laborers ($N1$), No. of employees ($N2$), Number of social responsibility activities in a year ($N3$), No. of products sold ($N4$), No. of operations involved ($N5$), No. of type of taxes ($N6$), No. of shipments involved ($N7$) and No. of other logistics ($N8$), while the outputs are the variables Volume of $CO_2$ generation ($VCO_2$), Energy consumption due to processes involved ($H_P$), Wastewater produced ($W_W$) and No. of products sold ($N4$) (% of sales) (in final case $VCO_2$ and $N_4$ are final output variables). Overall, there are 12 input variables and 4 output variables. Of these, the variables Volume of $CO_2$ generated ($VCO_2$), Energy consumption due to processes involved ($H_P$), Water used due to the processes involved ($W_P$), Wastewater produced ($W_W$), No. of employees ($N2$), Number of social responsibility activities in a year ($N3$), No. of other logistics ($N8$) have been drawn from the literature and No. of laborers ($N1$), No. of products sold ($N4$), No. of operations involved ($N5$), No. of type of taxes ($N6$), No. of transportations involved ($N7$) from the SMEs. Optimization of the free energy model leads to a reduced dimensional model involving only the 4 variables $N_4$, $N_5$, $N_7$, $V_{CO2}$, essentially implying information concentration within this 4-variable vector only. In the language of Machine Learning, we could have used formal dimensional reduction but what we find from our formalism is an equivalent representation which is at least equally powerful and more revealing from the perspective of system dynamics.

As an independent test study, we choose data from an anonymous Indian SME to test the robustness of the model. Two cases have been explored, separately in constrained and unconstrained environments, respectively for 3- and 5-year time spans. Results from case 1 show that carbon footprint ignorance may be less consequential for a short period in a constrained environment, but it will eventually drive bankruptcy. The carbon emission increase in this case represents negative growth that is environmentally unsustainable. Advance strategizing toward carbon footprint reduction led to an improved performance. This is a major testament of the strength of this model. Case 2 offers a scenario where sales reinforce carbon reduction targets leading to economic arbitrage where the SME will be operating in a break-even state. The carbon footprint is reduced to a setup target and re-strategizing offers direction towards sustainable production and consumption. This case also suggests that the time of action is independent of the cost function minimum rather it is a choice that remains with the SME. For a preliminary verification of the results, we conducted some elementary t-tests leading to p-values lying within $2.26 \times 10^{-8}$ - $1.92 \times 10^{-12}$. We used product sales fractions between 0.3 to 0.8 with number of operations involved between 70 to 85 for the above results. Detailed statistical validation is beyond the scope of this work though.

The two test cases can be separately solved using our own MATLAB codes within a few minutes, ensuring high computational efficiency of the algorithm proposed, leading towards a sustainable computing environment. The results clearly indicate that right strategy and the correct time of implementation play major role in business strategies, particularly in optimizing green production and profitability priorities. Our model is robust enough to handle diverse streamlined situations, e.g., SME and beyond, a marked improvement on existing practices that are largely limited to objective decision making instead of constrained subjective inputs as our mathematical architecture entails. The findings address SDG 9, 11, 12, 13, 14 and 15. Future studies involving different data sets from waste management sectors are underway with more generic kernel structure contributed by machine learning techniques (neural network, SVM) with an aim to deliver a consummate decision-making toolbox.

**Acknowledgement**

The authors acknowledge partial financial supports from the Commonwealth Scholarships Commission (Reference: INCN-2018-52) and Aston University. Dr. Prasanta Kumar Dey, Aston Business School is acknowledged for his general advice on supply chain literature.

# ONLINE SUPPLIMENTARY MATERIAL

## Appendix 1

### (Analytic Hierarchy Process - AHP)

**Steps of AHP**

The Analytic Hierarchy Process, abbreviated AHP, was developed by T. L. Saaty [A1] who used this technique to derive the weight factors of the uncertainty variables in the cost function, structured through the 'free-energy' model (Eq. 6). The general steps of performing AHP is described below:

*Step 1*: Defining the problem and determining its goal and structuring the hierarchy from the top through the intermediate levels to the lowest level containing the list of alternatives (see Figures 3 & 4).

*Step 2*: Construction of a set of pair-wise comparison matrices (size n x n) for each of the lower levels with one matrix for each element in the level immediately above by using the relative scale measurement. The pair-wise comparisons are done in terms of which element dominates the other. For our case, n=15, the number of generic variables considered in the SCN cost function in Eq. (6).

*Step 3*: Picked in groups of two over a set of n terms, this produces n(n-1)/2 judgments to develop the set of matrices in step 2. Reciprocals are automatically assigned in each pair-wise comparison.

*Step 4*: Hierarchical synthesis is now used to weight the eigenvectors (normalized representation in the generalized vector space) by the weights of the criteria and the sum is taken over all weighted eigenvector entries corresponding to those in the next lower level of the hierarchy. The AHP eigenvalues represent the normalized weights of the respective quantities, more in the mold of PCA.

*Step 5*: The consistency ratio (CR) confirms the reliability of the pairwise comparisons and it is determined as follows

$$\text{CR} = \frac{\lambda max - n}{n-1} \frac{CI}{RandomConsistency} \text{ CI/RI} \tag{S1}$$

Here consistency index CI = ($\lambda_{max}$ – n) / (n –1), where $\lambda_{max}$ is the maximum average value and n is the matrix size. The random consistency index (RI) depends on the value of n. CR is acceptable, if it is ≤ 0.10 for consistency. Beyond this value, the judgment matrix gets inconsistent. To obtain a consistent matrix, judgments should be reviewed and improved.

*Step 6*: Steps 2-5 are performed for all levels (3 levels as depicted in Figures 3 and 4) in the hierarchy.

**Description of the Layered AHP**

The goal of this AHP is to prioritize the alternatives from the perspective of Uncertainty. The criteria cluster consists of four nodes which are basically the four types of uncertainties considered for the study – Environmental, Economic, Social and Demand. The first layer of the alternatives consists of $VCO_2$, $W_p$, $H_p$ and $W_w$ connected from the Environmental node; $N_1$, $N_2$ and $N_3$ connected from the Social Node. The second layer of alternatives consists of $N_4$, $N_5$, $N_6$ and g connected from the economical node; $N_7$, $N_8$ and M connected from the Demand node. In order to establish the interdependencies among the parameters, the nodes in the second layer have been connected to the associated variables in the first layer. $V_{CO2}$ is connected to L, $N_5$ and $N_7$; $W_P$ is connected to $N_5$ & $N_6$; $H_p$ is connected to L and $N_5$; $W_w$ is connected to L and $N_5$ and $N_3$ is interlinked with $N_4$ and $N_6$. This AHP helps to find out the co-efficient of interdependencies directly from the AHP ($a_1$, $b_2$, $c_3$ etc). The compound interdependencies have been taken as product of concerned co-efficient, e.g. value of $a_{23}$ = value of $a_2$ x value of $a_3$. The squared co-efficient has been taken as square root of the concerned co-efficient. For example, the value of $a'_3$ = square root of $a_3$.

**Connections of AHP 1**

**Table S1: Node connection of AHP 1**

| Environmental ($S_1$) | Social ($S_2$) | Demand ($S_3$) | Economic ($S_4$) |
|---|---|---|---|
| Volume of $CO_2$ generated ($V_{CO2}$) | No. of labors ($N_1$) | No. of shipments ($N_7$) | No. of products sold ($N_4$) |
| Energy Consumption due to processes involved ($H_P$) | No. of employees ($N_2$) | No. of other logistics ($N_8$) | No. of operations involved ($N_5$) |
| Water used due to the processes involved ($W_P$) | No. of social responsibility activities in a year ($N_3$) | Miscellaneous cost ($M_1$) | No. of type of taxes ($N_6$) |
| Wastewater produced in the whole process ($W_w$) | | | Disaster management fund (g) |
| Legislative Costs (L) | | | |

**Connections of AHP 2**

**Table S2: Node connection of AHP 2 (for deriving the inter-dependency)**

| Category of Nodes in Layer 1 | Node in Layer 1 | Node Connected in Layer 2 | Category of Nodes in Layer 2 |
|---|---|---|---|
| Environment | Volume of $CO_2$ generated ($V_{CO2}$) | No. of shipments ($N_7$) | Demand |
| | | No. of operations involved ($N_5$) | Economic |
| | | Legislative Costs (L) | Environment |
| Environment | Energy Consumption due to processes involved ($H_P$) | No. of operations involved ($N_5$) | Economic |
| | | Legislative Costs (L) | Environment |
| Environment | Water used due to the processes involved ($W_P$) | No. of operations involved ($N_5$) | Economic |
| | | No. of type of taxes ($N_6$) | Economic |
| Environment | Wastewater produced in the whole process ($W_w$) | No. of operations involved ($N_5$) | Economic |
| | | Legislative Costs (L) | Environment |
| Social | No. of social responsibility activities in a year ($N_3$) | No. of products sold ($N_4$) | Economic |
| | | No. of operations involved ($N_5$) | Economic |

**Numerical Values Used for Calculation**

**Table S3: Values Derived using AHP used for calculation**

| Quantity | Value | Quantity | Value |
|---|---|---|---|
| $A_1$ | 0.0797 | $a_1$ | 0.10473 |
| $A_2$ | 0.0178 | $a_2$ | 0.25828 |
| $A_3$ | 0.0488 | $a_3$ | 0.63699 |
| $A_4$ | 0.0586 | $a_{12}$ | 0.02705 |
| $A_5$ | 0.0083 | $a_{23}$ | 0.164522 |
| $A_6$ | 0.0115 | $a_{31}$ | 0.066712 |
| $A_7$ | 0.0115 | $a'_1$ | 0.32362 |
| $A_8$ | 0.0573 | $a'_2$ | 0.508213 |
| $A_9$ | 0.1252 | $a'_3$ | 0.798117 |
| $A_{10}$ | 0.0963 | $b_1$ | 0.75 |
| $A_{11}$ | 0.0386 | $b'_1$ | 0.866025 |
| $A_{12}$ | 0.022 | $c_1$ | 0.5 |
| $A_{13}$ | 0.0773 | $c_2$ | 0.5 |
| $A_{14}$ | 0.1567 | $c_3$ | 0.25 |
| $A_{15}$ | 0.1906 | $c'_1$ | 0.707107 |
| $\varepsilon_1$ | 0.42458 | $c'_2$ | 0.707107 |
| $\varepsilon_2$ | 0.28198 | $d_1$ | 0.5 |
| $\varepsilon_3$ | 0.2132 | $d_2$ | 0.5 |
| $\varepsilon_4$ | 0.08024 | $d_3$ | 0.25 |
| $\lambda_1$ | 0.42458 | $d'_1$ | 0.707107 |
| $\lambda_2$ | 0.28198 | $d'_2$ | 0.707107 |
| $\lambda_3$ | 0.2132 | $\beta_1$ | 0.5 |
| $\lambda_4$ | 0.08024 | $\beta_2$ | 0.5 |
| $\alpha_1$ | 0.25 | $\beta_{12}$ | 0.25 |
| $\alpha_2$ | 0.75 | $\beta'_1$ | 0.707107 |
| $\alpha_{12}$ | 0.1875 | $\beta'_2$ | 0.707107 |
| $\alpha'_1$ | 0.5 | $\gamma$ | 0.1361 |
| $\alpha'_2$ | 0.866025 | | |

**Table S4: Data partially modified from an SME**

| Quantity | Value | Quantity | Value |
|---|---|---|---|
| $f_1$ | 10000 | $N_1$ | 42 |
| $f_2$ | 0 | $N_2$ | 9 |
| $f_3$ | 72000 | $N_3$ | 5 |
| $f_4$ | 60000 | $N_4$ | 15634 |
| $f_5$ | 108000 | $N_5$ | 76 |
| $f_6$ | 200000 | $N_6$ | 8 |
| $f_7$ | 160000 | $N_7$ | 2 |
| $f_8$ | 4500 | $N_8$ | 1 |
| $f_9$ | 0 | y | 0 |
| $f_{10}$ | 180000 | M | 166000 |
| $f_{11}$ | 0 | g | 25000 |
| u (scale factor) | 1/3000 | L | 17000 |

**Appendix 2**

**(The Unconstrained Model)**

A supply chain cost function is defined through a free energy functional in physics (mechanics) [A23]. This effectively represents the combination of input and output variables controlling system dynamics, often including constraints (as Lagrange multipliers). Since an 'energy' represents the functional component of a system, a free energy model is often referred to as a free energy functional. Our SCN model is represented a suitably weighted free energy functional where the weights define the relative importance of that variable with respect to others in the dynamical evolution of the system, information ingrained from SMEs.

*Generic Free energy F with weight factors*

$$F = \sum_i A_1 V_{CO_2} f_{1_i} + \sum_i A_2 H_{P_i} f_{2_i} + \sum_i A_3 W_{P_i} f_{3_i} + \sum_i A_4 W_{w_i} y_i + \sum_i A_5 L_i + \sum_i A_6 N_{1_i} f_{4_i}$$

$$+ \sum_i A_7 N_{2_i} f_{5_i} + \sum_i A_8 N_{3_i} f_{6_i} + + \sum_i A_9 N_{4_i} f_{7_i} - \sum_i A_{10} N_{5_i} f_{8_i} - \sum_i A_{11} f_{9_i} g_i - \sum_i A_{12} T_i N_{6_i}$$

$$+ \sum_i A_{14} f_{11_i} N_{8_i} + \sum_i A_{15} M_i$$

In this case, the index 'i' represents the number of operational days (1 < i < 300). This can be tailored to other values as per requirement.

**Detailed Mathematical Model**

The following is the cost function considering all the parameters and interdependencies.

$$F = \epsilon_1 \left( \sum_{i=1}^{300} A_1 V_{CO_{2_i}} f_{1_i} + \sum_{i=1}^{300} A_2 H_{P_i} f_{2_i} + \sum_{i=1}^{300} A_3 W_{P_i} f_{3_i} + \sum_{i=1}^{300} A_4 W_{w_i} y_i + \sum_{i=1}^{300} A_5 L_i \right)$$

$$+ \epsilon_2 \left( \sum_{i=1}^{300} A_6 N_{1_i} f_{4_i} + \sum_{i=1}^{300} A_7 N_{2_i} f_{5_i} + \sum_{i=1}^{300} A_8 N_{3_i} f_{6_i} \right)$$

$$+ \epsilon_3 \left( \sum_{i=1}^{300} A_9 N_{4_i} f_{7_i} - \sum_{i=1}^{300} A_{10} N_{5_i} f_{8_i} - \sum_{i=1}^{300} A_{11} g_i f_{9_i} - \sum_{i=1}^{300} A_{12} N_{6_i} T_i \right)$$

$$+ \epsilon_4 \left( \sum_{i=1}^{300} A_{13} N_{7_i} f_{10_i} + \sum_{i=1}^{300} A_{14} N_{8_i} f_{11_i} + \sum_{i=1}^{300} A_{15} M_i \right)$$

*Inter-dependency of the uncertainty variables*

The interdependencies has been estimated as two-degree polynomial and represented below

$$V_{CO_2} = a_1 L + a_2 N_5 + a_3 N_7 + a_{12} L N_5 + a_{23} N_5 N_7 + a_{31} L N_7 + a_1' L^2 + a_2' N_5^2 + a_3' N_7^2$$

$$W_P = W_P^O + b_1 N_5 + b_2 N_5^2$$

$$H_P = c_1 L + c_2 N_5 + c_{12} L N_5 + c_1' L^2 + c_2' N_5^2$$

$$W_w = d_1 L + d_2 N_5 + d_{12} L N_5 + d_1' L^2 + d_2' N_5^2$$

$$N_3 = \alpha_1 N_4 + \alpha_2 N_6 + \alpha_{12} N_4 N_6 + \alpha_1' N_4^2 + \alpha_2' N_5^2$$

$$N_4 = \beta_1 N_7 + \beta_2 N_8 + \beta_{12} N_7 N_8 + \beta_1' N_7^2 + \beta_2' N_8^2$$

$$N_7 = \gamma V_{CO_2}$$

**Table S5: Variables contributing to uncertainty along the supply chain network**

| Sl. | Symbols | Uncertainty Variable | Sources |
|---|---|---|---|
| 1 | $V_{CO2}$ | Volume of $CO_2$ generated | A2 |
| 2 | $H_p$ | Energy Consumption due to processes involved | A3 |
| 3 | $W_p$ | Water used due to the processes involved | A4 |
| 4 | $W_w$ | Wastewater produced in the whole process | A5 |
| 5 | $N_1$ | No. of labours | A21 |
| 6 | $N_2$ | No. of employees | A6 |
| 7 | $N_3$ | No. of social responsibility activities in a year | A7 |
| 8 | $N_4$ | No. of products sold | A8 |
| 9 | $N_5$ | No. of operations involved | A9 |
| 11 | $N_6$ | No. of type of taxes | A10 |
| 12 | $N_7$ | No. of shipments involved | A11 |
| 13 | $N_8$ | No. of other logistics | Synthetic data |
| 14 | $f_1$ | Unit cost for $CO_2$ recovery | A12 |
| 15 | $f_2$ | Unit cost for thermal pollution prevention | This study |

| 16 | f₃ | Unit cost for water used | A13 |
| 17 | f₄ | Wage of one labour | A21 |
| 18 | f₅ | Salary of one employee | A6 |
| 19 | f₆ | Average unit cost of social responsibility activity | A14 |
| 20 | f₇ | Unit revenue earned from product sold | A15 |
| 21 | f₈ | Unit cost of each operation | A16 |
| 22 | f₉ | Average cost for disaster management per day | A17 |
| 23 | f₁₀ | Unit cost for transportation | A20 |
| 24 | f₁₁ | Unit cost for other logistics | A19 |
| 25 | y | Unit cost for waste water treatment | A5 |
| 26 | M | Miscellaneous cost | This Study |
| 27 | L | Legislative Costs | A18 |
| 28 | T | Unit cost for Taxes | A16 |
| 29 | g | Disaster management fund | A17 |
| 30 | i | Iteration counter | A22 |

The process of optimization is presented in algorithmic form in Section 3.3. The cost function changes with the addition of 4 factors which are basically the constraints we imposed with the introduction of the Lagrange multipliers as shown in Eq. (9). The details of level and layer etc are now indicated in Figure 3 and 4. Additionally, the free energy model is described in section 3.2.2.

**First Derivatives**

$$\frac{\partial F}{\partial L} = \left[\varepsilon_1 A_1 a_1 + \varepsilon_1 A_2 f_2 c_1 + \varepsilon_1 A_4 y d_1 + \varepsilon_4 \gamma A_{13} f_{10} a_1 + \varepsilon_1 A_5\right] + \left[\varepsilon_1 A_1 a_{12} + \varepsilon_4 \gamma A_{13} f_{10} + \varepsilon_1 A_2 f_2 c_{12} + \varepsilon_1 A_4 y d_{12}\right] N_5$$

$$+ \left[\varepsilon_1 A_1 + \varepsilon_4 \gamma A_{13}\right] a_{31} N_7 + 2\left[\varepsilon_1 A_1 + \varepsilon_4 \gamma A_{13} f_{10} + \varepsilon_1 A_2 f_2 c_1' + \varepsilon_1 A_4 y d_1'\right] L$$

$$\frac{\partial F}{\partial N_1} = \varepsilon_2 A_6 f_4 \qquad \frac{\partial F}{\partial N_2} = \varepsilon_2 A_7 f_5$$

$$\frac{\partial F}{\partial N_3} = \varepsilon_2 A_8 f_6 + \varepsilon_3 \left[\frac{A_9 f_7}{\alpha_1 + \alpha_{12} N_6 + 2\alpha_1' N_4} - \frac{A_{12} T}{\alpha_2 + \alpha_{12} N_4 + 2\alpha_2' N_6}\right]$$

$$\frac{\partial F}{\partial N_4} = \varepsilon_3 A_9 f_7 + \varepsilon_2 A_8 f_6 \left(\alpha_1 + \alpha_{12} N_6 + 2\alpha_1' N_4\right) + \left[\frac{\varepsilon_4 A_{13} f_{10}}{\beta_1 + \beta_{12} N_8 + 2\beta_1' N_7} + \frac{\varepsilon_4 A_{14} f_{11}}{\beta_2 + \beta_{12} N_7 + 2\beta_2' N_8}\right]$$

$$\frac{\partial F}{\partial N_5} = \left[\varepsilon_1 A_1 f_1 a_2 + \varepsilon_4 \gamma A_{13} f_{10} a_2 + \varepsilon_1 A_2 f_2 c_2 + \varepsilon_1 A_4 y d_2 - \varepsilon_3 A_{10} f_8\right] +$$

$$\left[2N_5(\varepsilon_1 A_1 f_1 a_2' + \varepsilon_4 \gamma A_{13} f_{10} a_2' + \varepsilon_1 A_2 f_2 c_2' + \varepsilon_1 A_4 y d_2')\right]$$
$$+\left[\varepsilon_1 A_1 f_1 a_{23} + \varepsilon_4 \gamma A_{13} f_{10} a_{23}\right]N_7 + \left[\varepsilon_1 A_1 f_1 a_{12} + \varepsilon_1 A_2 f_2 c_{12} + \varepsilon_1 A_4 y d_{12} + \varepsilon_4 \gamma A_{13} f_{10} a_{12}\right]L$$

$$\frac{\partial V_{CO_2}}{\partial L} = a_1 + a_{12}N_5 + a_{31}N_7 + 2a_1'L$$

$$\frac{\partial W_P}{\partial N_5} = b_1 + 2b_1'N_5$$

$$\frac{\partial H_P}{\partial L} = c_1 + c_{12}N_5 + 2c_1'L$$

$$\frac{\partial H_P}{\partial N_5} = c_2 + c_{12}L + 2c_2'N_5$$

$$\frac{\partial W_w}{\partial L} = d_1 + d_{12}N_5 + 2d_1'L$$

$$\frac{\partial W_w}{\partial N_5} = d_2 + d_{12}L + 2d_2'N_5$$

$$\frac{\partial F}{\partial N_6} = -\varepsilon_3 A_{12}T + \varepsilon_2 A_8 f_6 \left(\alpha_2 + \alpha_{12}N_4 + 2\alpha_2'N_6\right)$$

$$\frac{\partial F}{\partial N_7} = \varepsilon_1 \left[A_1 f_1 \left(a_3 + a_{23}N_5 + a_{31}L + 2a_3'N_7\right)\right] + \varepsilon_3 \left[A_9 f_7 \left(\beta_1 + \beta_{12}N_8 + 2\beta_1'N_7\right)\right] + \varepsilon_4 A_{13} f_{10}$$

$$\frac{\partial F}{\partial N_8} = \varepsilon_3 A_9 f_7 \left[\left(\beta_2 + \beta_{12}N_7 + 2\beta_2'N_8\right)\right] + \varepsilon_4 A_{14} f_{11}$$

$$\frac{\partial F}{\partial g} = -\varepsilon_3 A_{11} f_9 \qquad \frac{\partial F}{\partial M} = \varepsilon_4 A_{15}$$

**Second derivatives**

$$\frac{\partial^2 F}{\partial V_{CO_2}^2} = \frac{1}{a_2 + a_{23}N_7 + a_{12}L + 2a_2'N_5}\left(\frac{\partial^2 F}{\partial V_{CO_2} \partial N_5}\right) + \frac{1}{a_1 + a_{12}N_5 + a_{31}N_7 + 2a_1'L}\left(\frac{\partial^2 F}{\partial V_{CO_2} \partial N_7}\right)$$

$$\frac{\partial^2 F}{\partial N_3^2} = \frac{1}{\alpha_1 + \alpha_{12}N_6 + 2\alpha_1'N_4} \times$$

$$\begin{pmatrix} \varepsilon_3 A_9 f_7 + \varepsilon_2 A_8 f_6 \left(\alpha_1 + \alpha_{12}N_6 + 2\alpha_1'N_4\right) \\ + \left[ \dfrac{\varepsilon_4 A_{13} f_{10}}{\beta_1 + \beta_{12}N_8 + 2\beta_1'N_7} + \dfrac{\varepsilon_4 A_{14} f_{11}}{\beta_2 + \beta_{12}N_7 + 2\beta_2'N_8} \right] \end{pmatrix}$$

$$+ \frac{1}{\alpha_2 + \alpha_{12}N_4 + 2\alpha_2'N_6}\left(\varepsilon_3 A_{12}T + \varepsilon_2 A_8 f_6 \left(\alpha_2 + \alpha_{12}N_4 + 2\alpha_2'N_6\right)\right)$$

$$\frac{\partial^2 F}{\partial N_4^2} = \left(\alpha_1 + \alpha_{12}N_6 + 2\alpha_1'N_4\right) \times$$

$$\left(\varepsilon_2 A_8 f_6 + \varepsilon_3 \left[ \frac{A_9 f_7}{\alpha_1 + \alpha_{12}N_6 + 2\alpha_1'N_4} - \frac{A_{12}T}{\alpha_2 + \alpha_{12}N_4 + 2\alpha_2'N_6} \right]\right)$$

$$+ \frac{1}{\beta_1 + \beta_{12}N_8 + 2\beta_1'N_7}\begin{pmatrix} \varepsilon_1 \left[ A_1 f_1 \left(a_3 + a_{23}N_5 + a_{31}L + 2a_3'N_7\right) \right] \\ +\varepsilon_3 \left[ A_9 f_7 \left(\beta_1 + \beta_{12}N_8 + 2\beta_1'N_7\right) \right] + \varepsilon_4 A_{13} f_{10} \end{pmatrix}$$

$$+ \frac{1}{\beta_2 + \beta_{12}N_7 + 2\beta_2'N_8}\left(\varepsilon_3 A_9 f_7 \left[\left(\beta_2 + \beta_{12}N_7 + 2\beta_2'N_8\right)\right] + \varepsilon_4 A_{14} f_{11}\right)$$

$$\frac{\partial^2 F}{\partial N_5^2} = 2\left(\varepsilon_1 A_1 f_1 a_2' + \gamma \varepsilon_4 A_{13} f_{10} a_2' + \varepsilon_1 A_2 f_2 c_2' + \varepsilon_1 A_4 y d_2'\right)$$

$$+ \gamma\left(a_2 + a_{23}N_7 + a_{12}L + 2a_2'N_5\right)\left[\gamma \varepsilon_4 A_{13} f_{10} a_{23} + \varepsilon_1 A_1 f_1 a_{23}\right]$$

$$\frac{\partial^2 F}{\partial N_7^2} = \frac{\varepsilon_1 A_1 f_1 a_{23}}{\gamma\left(a_2 + a_{23}N_7 + a_{12}L + 2a_2'N_5\right)} + 2\left(\varepsilon_1 A_1 f_1 a_3' + \varepsilon_3 A_9 f_7 \beta_1'\right)$$

$$\frac{\partial^2 F}{\partial N_3 \partial N_7} = -\frac{2\alpha_1' \varepsilon_3 A_9 f_7 \left(\beta_1 + \beta_{12}N_8 + 2\beta_1'N_7\right)}{\left(\alpha_1 + \alpha_{12}N_6 + 2\alpha_1'N_4\right)}$$

$$+ \frac{\alpha_{12} A_{12} T \left(\beta_1 + \beta_{12}N_8 + 2\beta_1'N_7\right)}{\left(\alpha_2 + \alpha_{12}N_4 + 2\alpha_2'N_6\right)^2}$$

$$\frac{\partial^2 F}{\partial N_3 \partial V_{CO_2}} = \left(-\frac{\gamma 2\alpha_1' \varepsilon_3 A_9 f_7 \left(\beta_1 + \beta_{12}N_8 + 2\beta_1'N_7\right)}{\left(\alpha_1 + \alpha_{12}N_6 + 2\alpha_1'N_4\right)} + \frac{\gamma \alpha_{12} A_{12} T \left(\beta_1 + \beta_{12}N_8 + 2\beta_1'N_7\right)}{\left(\alpha_2 + \alpha_{12}N_4 + 2\alpha_2'N_6\right)^2}\right)$$

$$\frac{\partial^2 F}{\partial N_4 \partial N_7} = \left(\gamma\left(a_1 + a_{12}N_5 + a_{31}N_7 + 2a_1'L\right)\right) \times$$

$$\left[\frac{2\beta_1'\varepsilon_4 A_{13} f_{10}}{\left(\beta_1 + \beta_{12}N_8 + 2\beta_1'N_7\right)^2} - \frac{\beta_{12}\varepsilon_4 A_{14} f_{11}}{\left(\beta_2 + \beta_{12}N_7 + 2\beta_2'N_8\right)^2}\right]$$

$$+2\varepsilon_2 A_8 f_6 \alpha_1' \left(\beta_1 + \beta_{12}N_8 + 2\beta_1'N_7\right)$$

$$\frac{\partial^2 F}{\partial N_4 \partial V_{CO_2}} = \frac{1}{\left(a_3 + a_{23}N_5 + a_{31}L + 2a_3'N_7\right)} \cdot \left(\frac{\partial^2 F}{\partial N_4 \partial N_7}\right)$$

$$\frac{\partial^2 F}{\partial N_5 \partial V_{CO_2}} = \left(\frac{1}{\left(a_1 + a_{12}N_5 + a_{31}N_7 + 2a_1'L\right)}\right) \times$$

$$\left(\begin{array}{c}\left(\varepsilon_1 A_1 a_{12} + \gamma\varepsilon_4 A_{13} f_{10} + \varepsilon_1 A_2 f_2 c_{12} + \varepsilon_1 A_4 y d_{12}\right) \\ +\gamma a_{13}\left(\varepsilon_1 A_1 + \varepsilon_4 A_{13}\gamma\right)\left(a_1 + a_{12}N_5 + a_{31}N_7 + 2a_1'L\right)\end{array}\right)$$

$$+\left(\frac{1}{\left(a_2 + a_{23}N_7 + a_{12}L + 2a_2'N_5\right)}\right) \times$$

$$\left(\begin{array}{c}2\left(\varepsilon_1 A_1 f_1 a_2' + \gamma\varepsilon_4 A_{13} f_{10} a_2' + \varepsilon_1 A_2 f_2 c_2' + \varepsilon_1 A_4 y d_2'\right) \\ +\gamma\left(a_2 + a_{23}N_7 + a_{12}L + 2a_2'N_5\right)\left[\gamma\varepsilon_4 A_{13} f_{10} a_{23} + \varepsilon_1 A_1 f_1 a_{23}\right]\end{array}\right)$$

$$\frac{\partial^2 F}{\partial N_7 \partial V_{CO_2}} = \frac{1}{\left(a_2 + a_{23}N_7 + a_{12}L + 2a_2'N_5\right)} \cdot \left(\frac{\partial^2 F}{\partial N_5 \partial N_7}\right)$$

$$+\frac{1}{\left(a_3 + a_{23}N_5 + a_{31}L + 2a_3'N_7\right)} \cdot \left(\frac{\partial^2 F}{\partial N_5 \partial N_7}\right)$$

$$\frac{\partial^2 F}{\partial N_5 \partial N_7} = \left(\frac{2\varepsilon_1 A_1 f_1 a_2' + \varepsilon_4 \gamma A_{13} f_{10} a_2' + 2\varepsilon_1 A_2 f_2 c_2' + 2A_4 y d_2'}{\gamma\left(a_1 + a_{12}N_5 + a_{31}N_7 + 2a_1'L\right)}\right) + \left[\varepsilon_1 A_1 f_1 a_{23} + \varepsilon_4 A_{13} f_{10} a_{23}\right]$$

$$+\left(\frac{\varepsilon_1 A_1 f_1 a_{12} + \varepsilon_1 A_2 f_2 c_{12} + \varepsilon_1 A_4 y d_{12} + \varepsilon_4 A_{13} f_{10} a_{12}}{\gamma\left(a_1 + a_{12}N_5 + a_{31}N_7 + 2a_1'L\right)}\right)$$

**Second derivatives in terms of the Lagrangian:**

$$\frac{\partial^2 \mathcal{L}}{\partial V_{CO_2}^2} = \frac{1}{a_2 + a_{23}N_7 + a_{12}L + 2a_2 N_5} \cdot \left( \frac{\partial^2 \mathcal{L}}{\partial V_{CO_2} \partial N_5} \right) + \frac{1}{a_3 + a_{23}N_5 + a_{13}L + 2a_3 N_7} \cdot \left( \frac{\partial^2 \mathcal{L}}{\partial V_{CO_2} \partial N_7} \right)$$

$$\frac{\partial^2 \mathcal{L}}{\partial V_{CO_2} \partial N_5} = \frac{1}{a_1 + a_{12}N_5 + a_{31}N_7 + 2a_1' L} \cdot \left[ \begin{array}{l} \left( \varepsilon_1 A_1 f_1 a_{12} + \gamma \varepsilon_4 A_{13} f_{10} + \varepsilon_1 A_2 f_2 c_{12} + \varepsilon_1 A_4 y d_{12} \right) \\ + \gamma a_{13} \left( \varepsilon_1 A_1 f_1 + \varepsilon_4 A_{13} \gamma f_{10} \right) \cdot \left( a_1 + a_{12}N_5 + a_{31}N_7 + 2a_1' L \right) \end{array} \right]$$

$$+ \frac{1}{a_2 + a_{32}N_7 + a_{21}LN_7 + 2a_2' N_5} \cdot \left[ \begin{array}{l} 2 \left( \varepsilon_1 A_1 f_1 a_2' + \gamma \varepsilon_4 A_{13} f_{10} a_2' + \varepsilon_1 A_2 f_2 c_2' + \varepsilon_1 A_4 y d_2 \right) \\ + \gamma \left( a_2 + a_{32}N_7 + a_{21}LN_7 + 2a_2' N_5 \right) \cdot \left( \gamma \varepsilon_4 A_{13} f_{10} a_{23} + \varepsilon_1 A_1 f_1 a_{23} + 2\varepsilon_1 A_1 f_1 b_2 \right) \end{array} \right]$$

$$\frac{\partial^2 \mathcal{L}}{\partial V_{CO_2} \partial N_7} = \frac{1}{a_2 + a_{32}N_7 + a_{21}L + 2a_2' N_5} \cdot \left( \frac{\partial^2 \mathcal{L}}{\partial N_5 \partial N_7} \right) + \frac{1}{a_3 + a_{23}N_5 + a_{31}L + 2a_3' N_7} \cdot \left( \frac{\partial^2 \mathcal{L}}{\partial N_7^2} \right)$$

$$\frac{\partial^2 \mathcal{L}}{\partial N_5 \partial N_7} = \left( \frac{2\varepsilon_1 A_1 f_1 a_2' + 2\gamma \varepsilon_4 A_{13} f_{10} a_2' + 2\varepsilon_1 A_2 f_2 c_2' + 2A_4 y d_2}{\gamma \left( a_1 + a_{12}N_5 + a_{31}N_7 + 2a_1' L \right)} \right) + \varepsilon_1 A_1 f_1 a_{13} + \varepsilon_4 A_{13} f_{10} a_{13}$$

$$+ \left( \frac{\varepsilon_1 A_1 f_1 a_{12} + \varepsilon_1 A_2 f_2 c_{12} + \varepsilon_1 A_4 y d_{12} + \gamma \varepsilon_{13} f_{10}}{\gamma \left( a_1 + a_{12}N_5 + a_{31}N_7 + 2a_1' L \right)} \right)$$

$$\frac{\partial^2 \mathcal{L}}{\partial N_7^2} = \left( \frac{\varepsilon_1 A_1 f_1 a_{23}}{\gamma \left( a_2 + a_{32}N_7 + a_{21}LN_7 + 2a_2' N_5 \right)} \right) + 2\varepsilon_1 A_1 f_1 a_3' + 2\varepsilon_3 A_9 f_7 \beta_1'$$

$$\frac{\partial^2 \mathcal{L}}{\partial V_{CO_2} \partial N_7} = \frac{1}{a_2 + a_{23}N_7 + a_{12}L + 2a_2' N_5} \cdot \left[ \begin{array}{l} \left( \dfrac{2\varepsilon_1 A_1 f_1 a_2' + 2\gamma \varepsilon_4 A_{13} f_{10} a_2' + 2\varepsilon_1 A_2 f_2 c_2' + 2A_4 y d_2}{\gamma \left( a_1 + a_{12}N_5 + a_{31}N_7 + 2a_1' L \right)} \right) + \varepsilon_1 A_1 f_1 a_{13} + \varepsilon_4 A_{13} f_{10} a_{13} \\ + \left( \dfrac{\varepsilon_1 A_1 f_1 a_{12} + \varepsilon_1 A_2 f_2 c_{12} + \varepsilon_1 A_4 y d_{12} + \gamma \varepsilon_{13} f_{10}}{\gamma \left( a_1 + a_{12}N_5 + a_{31}N_7 + 2a_1' L \right)} \right) \end{array} \right]$$

$$+ \frac{1}{a_3 + a_{23}N_5 + a_{31}L + 2a_3' N_7} \cdot \left[ \left( \frac{\varepsilon_1 A_1 f_1 a_{23}}{\gamma \left( a_2 + a_{32}N_7 + a_{21}LN_7 + 2a_2' N_5 \right)} \right) + 2\varepsilon_1 A_1 f_1 a_3' + 2\varepsilon_3 A_9 f_7 \beta_1' \right]$$

$$\frac{\partial^2 \mathcal{L}}{\partial V_{co_2}^2} = \frac{1}{a_2 + a_{23}N_7 + a_{12}L + 2a_2'N_5} \left[ \begin{array}{l} \dfrac{1}{a_1 + a_{12}N_5 + a_{31}N_7 + 2a_1'L} \cdot \left\{ \begin{array}{l} (\varepsilon_1 A_1 f_1 a_{12} + \gamma \varepsilon_4 A_{13} f_{10} + \varepsilon_1 A_2 f_2 c_{12} + \varepsilon_1 A_4 y d_{12}) \\ + \gamma a_{13} (\varepsilon_1 A_1 f_1 + \varepsilon_4 A_{13} \gamma f_{10}) \cdot (a_1 + a_{12}N_5 + a_{31}N_7 + 2a_1'L) \end{array} \right\} \\ + \dfrac{1}{a_2 + a_{23}N_7 + a_{12}LN_7 + 2a_2'N_5} \cdot \left\{ \begin{array}{l} 2(\varepsilon_1 A_1 f_1 a_2' + \gamma \varepsilon_4 A_{13} f_{10} a_2' + \varepsilon_1 A_2 f_2 c_2' + \varepsilon_1 A_4 y d_2') \\ + \gamma (a_2 + a_{32}N_7 + a_{21}LN_7 + 2a_2'N_5) \cdot (\gamma \varepsilon_4 A_{13} f_{10} a_{23} + \varepsilon_1 A_1 f_1 a_{23} + 2\varepsilon_1 A_1 f_1 b_2) \end{array} \right\} \end{array} \right]$$

$$+ \frac{1}{a_3 + a_{23}N_5 + a_{13}L + 2a_3'N_7} \cdot \left[ \begin{array}{l} \dfrac{1}{a_2 + a_{23}N_7 + a_{12}L + 2a_2'N_5} \cdot \left\{ \begin{array}{l} \left( \dfrac{2\varepsilon_1 A_1 f_1 a_2' + 2\gamma \varepsilon_4 A_{13} f_{10} a_2' + 2\varepsilon_1 A_2 f_2 c_2' + 2A_4 y d_2'}{\gamma (a_1 + a_{12}N_5 + a_{31}N_7 + 2a_1'L)} \right) + \varepsilon_1 A_1 f_1 a_{13} + \varepsilon_4 A_{13} f_{10} a_{13} \\ + \left( \dfrac{\varepsilon_1 A_1 f_1 a_{12} + \varepsilon_1 A_2 f_2 c_{12} + \varepsilon_1 A_4 y d_{12} + \gamma \varepsilon_{13} f_{10}}{\gamma (a_1 + a_{12}N_5 + a_{31}N_7 + 2a_1'L)} \right) \end{array} \right\} \\ + \dfrac{1}{a_3 + a_{23}N_5 + a_{31}L + 2a_3'N_7} \cdot \left\{ \left( \dfrac{\varepsilon_1 A_1 f_1 a_{23}}{\gamma (a_2 + a_{23}N_7 + a_{12}L + 2a_2'N_5)} \right) + 2\varepsilon_1 A_1 f_1 a_3' + 2\varepsilon_3 A_9 f_7 \beta_1' \right\} \end{array} \right]$$

$$\frac{\partial^2 \mathcal{L}}{\partial N_5^2} = 2(\varepsilon_1 A_1 f_1 a_2' + \gamma \varepsilon_4 A_{13} f_{10} a_2' + \varepsilon_1 A_2 f_2 c_2' + A_4 y d_2') + \gamma (a_2 + a_{23}N_7 + a_{12}L + 2a_2'N_5) \cdot (\gamma \varepsilon_4 A_{13} f_{10} a_{23} + \varepsilon_1 A_1 f_1 a_{23}) + 2\varepsilon_1 A_3 f_3 b_2$$

$$\frac{\partial^2 \mathcal{L}}{\partial N_4^2} = \left(\alpha_1 + \alpha_{12}N_6 + 2\alpha_1' N_4\right) \cdot \left\{ \varepsilon_2 A_8 f_6 + \varepsilon_3 \cdot \left( \frac{A_9 f_7}{\alpha_1 + \alpha_{12}N_6 + 2\alpha_1' N_4} - \frac{A_{12}T}{\alpha_2 + \alpha_{12}N_4 + 2\alpha_2' N_6} \right) \right\}$$

$$+ \frac{1}{\beta_1 + \beta_{12}N_8 + 2\beta_1' N_7} \cdot \left[ \varepsilon_1 \left\{ A_1 f_1 \left( a_3 + a_{23}N_5 + a_{31}L + 2a_3' N_7 \right) \right\} + \varepsilon_3 \left\{ A_9 f_7 \left( \beta_1 + \beta_{12}N_8 + 2\beta_1' N_7 \right) \right\} + \varepsilon_4 A_{13} f_{10} \right]$$

$$+ \frac{1}{\beta_2 + \beta_{12}N_7 + 2\beta_2' N_8} \cdot \left[ \varepsilon_3 \left\{ A_9 f_7 \left( \beta_1 + \beta_{12}N_8 + 2\beta_1' N_7 \right) \right\} + \varepsilon_4 A_{14} f_{11} \right]$$

$$\frac{\partial^2 \mathcal{L}}{\partial N_4 \partial V_{CO_2}} = \frac{1}{a_2 + a_{23}N_7 + a_{12}L + 2a_2' N_5} \cdot \left( \frac{\partial \mathcal{L}^2}{\partial N_4 \partial N_7} \right)$$

$$\frac{\partial^2 \mathcal{L}}{\partial N_4 \partial N_7} = \gamma \left( a_1 + a_{12}N_5 + a_{31}N_7 + 2a_1' L \right) \cdot \left( \frac{2\beta_1' \varepsilon_4 A_{13} f_{10}}{\beta_1 + \beta_{12}N_8 + 2\beta_1' N_7} - \frac{\beta_{12} \varepsilon_4 A_{14} f_{11}}{\beta_2 + \beta_{12}N_7 + 2\beta_2' N_8} \right)$$

$$+ 2\varepsilon_2 A_8 f_6 \alpha_1' \cdot \left( \beta_1 + \beta_{12}N_8 + 2\beta_1' N_7 \right)$$

$$\frac{\partial^2 \mathcal{L}}{\partial N_4 \partial V_{CO_2}} = \frac{1}{a_2 + a_{23}N_7 + a_{12}L + 2a_2' N_5} \cdot \left[ \begin{array}{l} \gamma \left( a_1 + a_{12}N_5 + a_{31}N_7 + 2a_1' L \right) \cdot \left( \dfrac{2\beta_1' \varepsilon_4 A_{13} f_{10}}{\beta_1 + \beta_{12}N_8 + 2\beta_1' N_7} - \dfrac{\beta_{12} \varepsilon_4 A_{14} f_{11}}{\beta_2 + \beta_{12}N_7 + 2\beta_2' N_8} \right) \\ + 2\varepsilon_2 A_8 f_6 \alpha_1' \cdot \left( \beta_1 + \beta_{12}N_8 + 2\beta_1' N_7 \right) \end{array} \right]$$

$$H = \begin{bmatrix}
\frac{\partial^2 F}{\partial V_{CO_2}^2} & \frac{\partial^2 F}{\partial V_{CO_2}\partial N_1} & \frac{\partial^2 F}{\partial V_{CO_2}\partial N_2} & \frac{\partial^2 F}{\partial V_{CO_2}\partial N_3} & \frac{\partial^2 F}{\partial V_{CO_2}\partial N_4} & \frac{\partial^2 F}{\partial V_{CO_2}\partial N_5} & \frac{\partial^2 F}{\partial V_{CO_2}\partial N_6} & \frac{\partial^2 F}{\partial V_{CO_2}\partial N_7} & \frac{\partial^2 F}{\partial V_{CO_2}\partial N_8} & \frac{\partial^2 F}{\partial V_{CO_2}\partial W_P} & \frac{\partial^2 F}{\partial V_{CO_2}\partial H_P} & \frac{\partial^2 F}{\partial V_{CO_2}\partial W_w} \\
\frac{\partial^2 F}{\partial N_1\partial V_{CO_2}} & \frac{\partial^2 F}{\partial N_1^2} & \frac{\partial^2 F}{\partial N_1\partial N_2} & \frac{\partial^2 F}{\partial N_1\partial N_3} & \frac{\partial^2 F}{\partial N_1\partial N_4} & \frac{\partial^2 F}{\partial N_1\partial N_5} & \frac{\partial^2 F}{\partial N_1\partial N_6} & \frac{\partial^2 F}{\partial N_1\partial N_7} & \frac{\partial^2 F}{\partial N_1\partial N_8} & \frac{\partial^2 F}{\partial N_1\partial W_P} & \frac{\partial^2 F}{\partial N_1\partial H_P} & \frac{\partial^2 F}{\partial N_1\partial W_w} \\
\frac{\partial^2 F}{\partial N_2\partial V_{CO_2}} & \frac{\partial^2 F}{\partial N_2\partial N_1} & \frac{\partial^2 F}{\partial N_2^2} & \frac{\partial^2 F}{\partial N_2\partial N_3} & \frac{\partial^2 F}{\partial N_2\partial N_4} & \frac{\partial^2 F}{\partial N_2\partial N_5} & \frac{\partial^2 F}{\partial N_2\partial N_6} & \frac{\partial^2 F}{\partial N_2\partial N_7} & \frac{\partial^2 F}{\partial N_2\partial N_8} & \frac{\partial^2 F}{\partial N_2\partial W_P} & \frac{\partial^2 F}{\partial N_2\partial H_P} & \frac{\partial^2 F}{\partial N_2\partial W_w} \\
\frac{\partial^2 F}{\partial N_3\partial V_{CO_2}} & \frac{\partial^2 F}{\partial N_3\partial N_1} & \frac{\partial^2 F}{\partial N_3\partial N_2} & \frac{\partial^2 F}{\partial N_3^2} & \frac{\partial^2 F}{\partial N_3\partial N_4} & \frac{\partial^2 F}{\partial N_3\partial N_5} & \frac{\partial^2 F}{\partial N_3\partial N_6} & \frac{\partial^2 F}{\partial N_3\partial N_7} & \frac{\partial^2 F}{\partial N_3\partial N_8} & \frac{\partial^2 F}{\partial N_3\partial W_P} & \frac{\partial^2 F}{\partial N_3\partial H_P} & \frac{\partial^2 F}{\partial N_3\partial W_w} \\
\frac{\partial^2 F}{\partial N_4\partial V_{CO_2}} & \frac{\partial^2 F}{\partial N_4\partial N_1} & \frac{\partial^2 F}{\partial N_4\partial N_2} & \frac{\partial^2 F}{\partial N_4\partial N_3} & \frac{\partial^2 F}{\partial N_4^2} & \frac{\partial^2 F}{\partial N_4\partial N_5} & \frac{\partial^2 F}{\partial N_4\partial N_6} & \frac{\partial^2 F}{\partial N_4\partial N_7} & \frac{\partial^2 F}{\partial N_4\partial N_8} & \frac{\partial^2 F}{\partial N_4\partial W_P} & \frac{\partial^2 F}{\partial N_4\partial H_P} & \frac{\partial^2 F}{\partial N_4\partial W_w} \\
\frac{\partial^2 F}{\partial N_5\partial V_{CO_2}} & \frac{\partial^2 F}{\partial N_5\partial N_1} & \frac{\partial^2 F}{\partial N_5\partial N_2} & \frac{\partial^2 F}{\partial N_5\partial N_3} & \frac{\partial^2 F}{\partial N_5\partial N_4} & \frac{\partial^2 F}{\partial N_5^2} & \frac{\partial^2 F}{\partial N_5\partial N_6} & \frac{\partial^2 F}{\partial N_5\partial N_7} & \frac{\partial^2 F}{\partial N_5\partial N_8} & \frac{\partial^2 F}{\partial N_5\partial W_P} & \frac{\partial^2 F}{\partial N_5\partial H_P} & \frac{\partial^2 F}{\partial N_5\partial W_w} \\
\frac{\partial^2 F}{\partial N_6\partial V_{CO_2}} & \frac{\partial^2 F}{\partial N_6\partial N_1} & \frac{\partial^2 F}{\partial N_6\partial N_2} & \frac{\partial^2 F}{\partial N_6\partial N_3} & \frac{\partial^2 F}{\partial N_6\partial N_4} & \frac{\partial^2 F}{\partial N_6\partial N_5} & \frac{\partial^2 F}{\partial N_6^2} & \frac{\partial^2 F}{\partial N_6\partial N_7} & \frac{\partial^2 F}{\partial N_6\partial N_8} & \frac{\partial^2 F}{\partial N_6\partial W_P} & \frac{\partial^2 F}{\partial N_6\partial H_P} & \frac{\partial^2 F}{\partial N_6\partial W_w} \\
\frac{\partial^2 F}{\partial N_7\partial V_{CO_2}} & \frac{\partial^2 F}{\partial N_7\partial N_1} & \frac{\partial^2 F}{\partial N_7\partial N_2} & \frac{\partial^2 F}{\partial N_7\partial N_3} & \frac{\partial^2 F}{\partial N_7\partial N_4} & \frac{\partial^2 F}{\partial N_7\partial N_5} & \frac{\partial^2 F}{\partial N_7\partial N_6} & \frac{\partial^2 F}{\partial N_7^2} & \frac{\partial^2 F}{\partial N_7\partial N_8} & \frac{\partial^2 F}{\partial N_7\partial W_P} & \frac{\partial^2 F}{\partial N_7\partial H_P} & \frac{\partial^2 F}{\partial N_7\partial W_w} \\
\frac{\partial^2 F}{\partial N_8\partial V_{CO_2}} & \frac{\partial^2 F}{\partial N_8\partial N_1} & \frac{\partial^2 F}{\partial N_8\partial N_2} & \frac{\partial^2 F}{\partial N_8\partial N_3} & \frac{\partial^2 F}{\partial N_8\partial N_4} & \frac{\partial^2 F}{\partial N_8\partial N_5} & \frac{\partial^2 F}{\partial N_8\partial N_6} & \frac{\partial^2 F}{\partial N_8\partial N_7} & \frac{\partial^2 F}{\partial N_8^2} & \frac{\partial^2 F}{\partial N_8\partial W_P} & \frac{\partial^2 F}{\partial N_8\partial H_P} & \frac{\partial^2 F}{\partial N_8\partial W_w} \\
\frac{\partial^2 F}{\partial W_P\partial V_{CO_2}} & \frac{\partial^2 F}{\partial W_P\partial N_1} & \frac{\partial^2 F}{\partial W_P\partial N_2} & \frac{\partial^2 F}{\partial W_P\partial N_3} & \frac{\partial^2 F}{\partial W_P\partial N_4} & \frac{\partial^2 F}{\partial W_P\partial N_5} & \frac{\partial^2 F}{\partial W_P\partial N_6} & \frac{\partial^2 F}{\partial W_P\partial N_7} & \frac{\partial^2 F}{\partial W_P\partial N_8} & \frac{\partial^2 F}{\partial W_P^2} & \frac{\partial^2 F}{\partial W_P\partial H_P} & \frac{\partial^2 F}{\partial W_P\partial W_w} \\
\frac{\partial^2 F}{\partial H_P\partial V_{CO_2}} & \frac{\partial^2 F}{\partial H_P\partial N_1} & \frac{\partial^2 F}{\partial H_P\partial N_2} & \frac{\partial^2 F}{\partial H_P\partial N_3} & \frac{\partial^2 F}{\partial H_P\partial N_4} & \frac{\partial^2 F}{\partial H_P\partial N_5} & \frac{\partial^2 F}{\partial H_P\partial N_6} & \frac{\partial^2 F}{\partial H_P\partial N_7} & \frac{\partial^2 F}{\partial H_P\partial N_8} & \frac{\partial^2 F}{\partial H_P\partial W_P} & \frac{\partial^2 F}{\partial H_P^2} & \frac{\partial^2 F}{\partial H_P\partial W_w} \\
\frac{\partial^2 F}{\partial W_w\partial V_{CO_2}} & \frac{\partial^2 F}{\partial W_w\partial N_1} & \frac{\partial^2 F}{\partial W_w\partial N_2} & \frac{\partial^2 F}{\partial W_w\partial N_3} & \frac{\partial^2 F}{\partial W_w\partial N_4} & \frac{\partial^2 F}{\partial W_w\partial N_5} & \frac{\partial^2 F}{\partial W_w\partial N_6} & \frac{\partial^2 F}{\partial W_w\partial N_7} & \frac{\partial^2 F}{\partial W_w\partial N_8} & \frac{\partial^2 F}{\partial W_w\partial W_P} & \frac{\partial^2 F}{\partial W_w\partial H_P} & \frac{\partial^2 F}{\partial W_w^2}
\end{bmatrix}$$

The above H-matrix leads to the constrained 4x4 Hessian as below (details in the main text):

$$H = \begin{bmatrix}
\frac{\partial^2 \mathcal{L}}{\partial V_{CO_2}^2} & 0 & 0 & 0 \\
\frac{\partial^2 \mathcal{L}}{\partial N_4\partial V_{CO_2}} & \frac{\partial^2 \mathcal{L}}{\partial N_4^2} & 0 & \frac{\partial^2 \mathcal{L}}{\partial N_4\partial N_7} \\
\frac{\partial^2 \mathcal{L}}{\partial N_5\partial V_{CO_2}} & 0 & \frac{\partial^2 \mathcal{L}}{\partial N_5^2} & \frac{\partial^2 \mathcal{L}}{\partial N_5\partial N_7} \\
\frac{\partial^2 \mathcal{L}}{\partial N_7\partial V_{CO_2}} & 0 & 0 & \frac{\partial^2 \mathcal{L}}{\partial N_7^2}
\end{bmatrix}$$

**The Euler-Lagrangian Structure**

*Unconstrained Model*

$$\delta\left(\frac{d}{dt}\begin{bmatrix}\frac{\partial F}{\partial N_4}\\ \frac{\partial F}{\partial N_5}\\ \frac{\partial F}{\partial N_3}\\ \frac{\partial F}{\partial N_1}\\ \frac{\partial F}{\partial N_2}\\ \frac{\partial F}{\partial N_6}\\ \frac{\partial F}{\partial N_7}\\ \frac{\partial F}{\partial N_8}\\ \frac{\partial F}{\partial V_{CO_2}}\\ \frac{\partial F}{\partial L}\end{bmatrix}\right) = \begin{bmatrix} x_{1,1} & 0 & 0 & 0 & 0 & 0 & x_{1,7} & x_{1,8} & 0 & 0 \\ 0 & x_{2,2} & 0 & 0 & 0 & 0 & x_{2,7} & 0 & 0 & x_{2,10} \\ x_{3,1} & 0 & x_{3,3} & 0 & 0 & x_{3,6} & 0 & 0 & 0 & 0 \\ 0 & 0 & 0 & 0 & 0 & 0 & 0 & 0 & 0 & 0 \\ 0 & 0 & 0 & 0 & 0 & 0 & 0 & 0 & 0 & 0 \\ x_{6,1} & 0 & 0 & 0 & 0 & x_{6,6} & 0 & 0 & 0 & 0 \\ 0 & x_{7,2} & 0 & 0 & 0 & 0 & x_{7,7} & x_{7,8} & 0 & x_{7,10} \\ 0 & 0 & 0 & 0 & 0 & 0 & x_{8,7} & x_{8,8} & 0 & 0 \\ 0 & x_{9,2} & 0 & 0 & 0 & x_{9,6} & 0 & 0 & x_{9,9} & x_{9,10} \\ 0 & x_{10,2} & 0 & 0 & 0 & 0 & x_{10,7} & 0 & 0 & x_{10,10}\end{bmatrix}\begin{bmatrix}\delta N_4\\ \delta N_5\\ \delta N_2\\ \delta N_1\\ \delta N_2\\ \delta N_6\\ \delta N_7\\ \delta N_8\\ \delta V_{CO_2}\\ \delta L\end{bmatrix}$$

This leads to the following after removing the zero rows and columns, which translates to the dynamical model

$$\delta\left(\frac{d}{dt}\begin{bmatrix}\frac{\partial F}{\partial N_4}\\ \frac{\partial F}{\partial N_5}\\ \frac{\partial F}{\partial N_7}\\ \frac{\partial F}{\partial V_{CO_2}}\end{bmatrix}\right) = \begin{bmatrix} m_{11} & 0 & m_{13} & 0 \\ 0 & m_{22} & m_{23} & 0 \\ 0 & m_{32} & m_{33} & 0 \\ 0 & m_{42} & 0 & m_{44}\end{bmatrix}\begin{bmatrix}\delta N_4\\ \delta N_5\\ \delta N_7\\ \delta V_{CO_2}\end{bmatrix}$$

$$\frac{d^2}{dt^2}\begin{bmatrix} \xi_1 \delta N_4 \\ \xi_2 \delta N_5 \\ \xi_3 \delta N_7 \\ \xi_4 \delta V_{CO_2} \end{bmatrix} = \begin{bmatrix} m_{11} & 0 & m_{13} & 0 \\ 0 & m_{22} & m_{23} & 0 \\ 0 & m_{32} & m_{33} & 0 \\ 0 & m_{42} & 0 & m_{44} \end{bmatrix} \begin{bmatrix} \delta N_4 \\ \delta N_5 \\ \delta N_7 \\ \delta V_{CO_2} \end{bmatrix}$$

*Constrained Model*

The reduced version the constrained model takes this form –

$$\delta \left( \frac{d}{dt} \begin{bmatrix} \frac{\partial L}{\partial N_4} \\ \frac{\partial L}{\partial N_5} \\ \frac{\partial L}{\partial N_7} \\ \frac{\partial L}{\partial V_{CO_2}} \end{bmatrix} \right) = \begin{bmatrix} k_{11} & 0 & k_{13} & 0 \\ 0 & k_{22} & k_{23} & 0 \\ 0 & k_{32} & k_{33} & 0 \\ 0 & k_{42} & 0 & k_{44} \end{bmatrix} \begin{bmatrix} \delta N_4 \\ \delta N_5 \\ \delta N_7 \\ \delta V_{CO_2} \end{bmatrix}$$

which translates to the following dynamical model

$$\frac{d^2}{dt^2}\begin{bmatrix} \psi_1 \delta N_4 \\ \psi_2 \delta N_5 \\ \psi_3 \delta N_7 \\ \psi_4 \delta V_{CO_2} \end{bmatrix} = \begin{bmatrix} k_{11} & 0 & k_{13} & 0 \\ 0 & k_{22} & k_{23} & 0 \\ 0 & k_{32} & k_{33} & 0 \\ 0 & k_{42} & 0 & k_{44} \end{bmatrix} \begin{bmatrix} \delta N_4 \\ \delta N_5 \\ \delta N_7 \\ \delta V_{CO_2} \end{bmatrix}$$

**Elements of Matrix m and k**

$M_{11} = K_{11} = 2\epsilon_2 A_8 f_6 \alpha'_1$

$M_{13} = K_{13} = -\epsilon_4 A_{13} f_{10} \dfrac{2\beta'_1}{(\beta_1 + \beta_{12} N_8^0 + 2\beta'_1 N_7^0)^2} - \epsilon_4 A_{14} f_{11} \dfrac{\beta_{12}}{(\beta_1 + \beta_{12} N_7^0 + 2\beta'_2 N_8^0)^2}$

$M_{22} = K_{22} = 2(\epsilon_1 A_1 f_1 a'_2 + \gamma \epsilon_4 A_{13} f_{10} a'_2 + \epsilon_1 A_2 f_2 c'_2 + \epsilon_1 A_4 y d'_2 + \epsilon_1 A_3 f_3 b_2)$

$M_{23} = K_{23} = (\epsilon_1 A_1 f_1 a_{23} + \gamma \epsilon_4 A_{13} f_{10} a_{23})$

$M_{32} = K_{32} = \epsilon_1 A_1 f_1 a_{23}$

$M_{33} = K_{33} = 2\epsilon_1 A_1 f_1 a'_3 + 2\epsilon_3 A_9 f_7 \beta'_1$

$M_{42} = K_{42} = \dfrac{-a_{12}}{(a_1 + a_{12} N_5 + a_{31} N_7 + 2a'_1 L)^2} [(\epsilon_1 A_1 f_1 a_{12} + \gamma \epsilon_4 A_{13} f_{10} + \epsilon_1 A_2 f_2 c_{12} + \epsilon_1 A_4 y d_{12}) N_5$
$\qquad + (\epsilon_1 A_1 f_1 a_{31} + \gamma \epsilon_4 A_{13} f_{10} a_{31}) N_7 + 2(\epsilon_1 A_1 f_1 a'_1 + \gamma \epsilon_4 A_{13} f_{10} a'_1 + \epsilon_1 A_2 f_2 c'_1 + \epsilon_1 A_4 y d'_1) L]$
$\qquad + \dfrac{1}{(a_1 + a_{12} N_5 + a_{31} N_7 + 2a'_1 L)}$

$$M_{44} = K_{44} = -\lambda_3 f_1$$

**Results of Unconstrained Optimization for VCO$_2$**

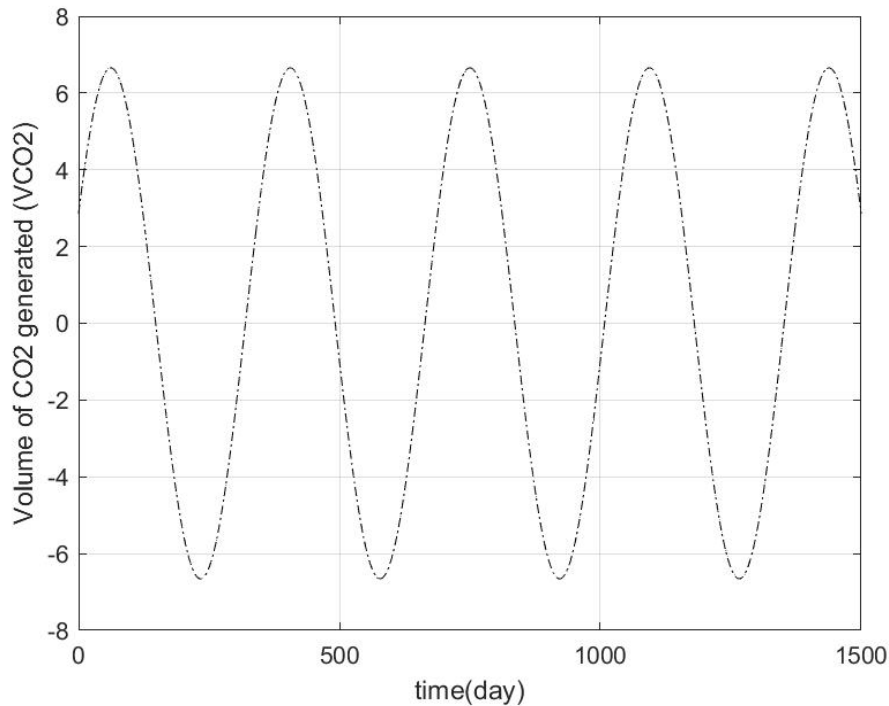

Figure S1: Time Dependence of VCO$_2$ in a constrained environment: 5-year time span

As mentioned in section 5.1, V$_{CO2}$ gives non-converging solutions. The time dependent plot of VCO$_2$ in an unconstrained environment is presented in Figure S1. The curve has an oscillatory profile which implies that in the unconstrained environment, the system is unstable without any control of carbon emission.

**Extended Bibliography**

v

43